\documentclass[10pt]{article}
\usepackage[utf8]{inputenc}
\usepackage{amsmath}
\usepackage{float}
\usepackage{amsfonts}

\usepackage{caption}
\usepackage{subcaption}

\usepackage[left=2cm,right=2cm,top=2cm,bottom=2cm]{geometry}
\usepackage{csquotes}

\usepackage[pdftex]{graphicx}
\usepackage{epsfig}
\usepackage{epstopdf}

\usepackage{amssymb}
\usepackage{dirtytalk}
\usepackage{xurl}
\usepackage[hidelinks,breaklinks]{hyperref}
\usepackage[nameinlink]{cleveref}
\usepackage{xcolor}
\hypersetup{
    colorlinks,
    linkcolor={red},
    citecolor={green!60!black},
    urlcolor={blue!80!black}
}

\usepackage[sorting=none,style=ieee] {biblatex}
\DeclareUnicodeCharacter{0301}{\`{e}}

\DeclareUnicodeCharacter{2265}{~}

\usepackage{relsize,exscale}
\usepackage{latexsym}

\title{\bf Accretion Phenomena of Different Kinds of Chaplygin Gas Models onto Kehagias-Sfetsos Black Hole
in Ho{\v{r}}ava-Lifshitz Gravity Scenario }
\author{\bf Puja Mukherjee$^1$\footnote{e-mail:{pmukherjee967@gmail.com}}~,~
Ujjal Debnath$^1$\footnote{e-mail: {ujjaldebnath@gmail.com}} ~and~
Anirudh Pradhan$^2$\footnote{e-mail :
{pradhan.anirudh@gmail.com}}\\
$^1$Department of Mathematics, Indian Institute of Engineering
Science\\ and Technology, Shibpur, Howrah-711103, India.\\
$^2$Centre for Cosmology, Astrophysics, and Space Science,\\ GLA
University, Mathura-281406, U.P., India. }

\date{\today}

\begin{document}


\maketitle


\begin{abstract}
 In the Ho\v{r}ava-Lifshitz gravity scenario, we have described the mass accretion process of a 4-dimensional Kehagias-Sfetsos black hole caused by some candidates of dark energy in detail. Firstly, we have obtained the equation of mass for the black hole with respect to the cosmic scale parameter `$a$' in the Ho{\v{r}}ava-Lifshitz gravity framework. Then, one by one, we have considered dark energies like Generalized Cosmic Chaplygin gas, Variable Modified Chaplygin gas, New Variable Modified Chaplygin gas, Modified Chaplygin-Jacobi gas, and Modified Chaplygin-Abel gas, respectively, and established their equation of mass in terms of the redshift function `z.' In addition, we have presented the results with graphical representations of the mass and redshift functions. Finally, we came to the conclusion that, due to the accretion of all of the dark energy candidates mentioned above, the mass of a 4-dimensional Kehagias-Sfetsos Black Hole will increase gradually in correspondence to the evolution of the universe.
\end{abstract}




\section{Introduction}\label{a}
The last century proved to be a golden era for cosmology, as all the pieces of evidence starting from observation of type Ia supernovae
\cite{perlmutter1998cosmology,garnavich1998constraints,bahcall1999cosmic,perlmutter1999supernova,filippenko1998results}, large-scale structure theory \cite{filippenko1998results,abazajian2004second,perlmutter1999measurements,perlmutter1998discovery,tegmark2004cosmological,perlmutter1997measurements}, Chandra X-ray observations \cite{allen2004constraints}, Planck observations \cite{adam2016planck}, observations regarding CMB i.e., cosmic microwave background radiations \cite{spergel2007three} and observation of Baryon acoustic oscillation (BAO) \cite{eisenstein1998baryonic,eisenstein2005detection} to WMAP prediction \cite{bennett2003first,spergel2003first,verde20022df} each of these indicated towards the confirmation of Edwin Hubble's discovery about the expansion of our universe. Not only that, these observations have also verified the theory of the accelerated rate of expansion of the universe. This theory of accelerated expansion unveiled another peculiar concept called ``Dark energy" (DE). DE is a strange kind of energy with negative pressure, violates the strong energy condition (i.e., $\rho+3p<0$), and leads to accelerated expansion of the universe. According to cosmologists, DE is the dominant energy of our universe and has an antigravity effect.

After finding its cosmic acceleration trait, cosmologists started to explore the models that can be considered in place of DE. Mainly, the categorization of DE depends upon the value of parameter ``$\omega$", popularly known as equation of state parameter. It's most common form is called cosmological constant or vacuum energy density, denoted by ``$\Lambda$", for the value of $\omega=-1$  \cite{padmanabhan2003cosmological,carroll1992cosmological,peebles2003cosmological,bianchi2010all,weinberg2000cosmological,copeland2006dynamics}, and the corresponding model is known as $\Lambda CDM$ model.Other than this static kind of DE, there are several dynamic kind of DEs like: quintessence \cite{caldwell1998cosmological,peebles1988cosmology,bamba2012dark,chen2015constraints,smer2017planck,wetterich1988cosmology}, K-essence \cite{armendariz2001essentials}, phantom energy \cite{caldwell2002phantom,elizalde2004late}, cosmic scalar fields which advances with potential and have fluctuations \cite{caldwell1998cosmological}, tachyonic field \cite{sen2002rolling,gibbons2002cosmological}, DBI-essence \cite{martin2008dbi,chakraborty2020investigating} etc. Again, to solve problems on fine-tuning and cosmic coincidence \cite{padmanabhan2003cosmological,copeland2006dynamics,weinberg1989cosmological}, we mostly make use of some scalar field models \cite{armendariz2000dynamical,sen2002tachyon,feng2005dark,guo2005cosmological,gasperini2001quintessence,gumjudpai2009generalized,martin2008dbi,wei2005hessence} acting like DE.\\

Modified gravity theories
\cite{dvali20004d,jacobson2001gravity,abdalla2005consistent,hovrava2009membranes,barrow2012some,meng2012einstein,nojiri2005modified,tsujikawa2010modified}
are another celebrated topic among researchers to fill in place of DE in connection with measuring the accelerated universe expansion. As time progresses, various research studies have been conducted on combining dark energy and dark matter to get a new kind of model that can help us measure the current acceleration rate of our universe. This gives us one more alternative of DE, called pure Chaplygin gas \cite{kamenshchik2001alternative,bilic2002unification,gorini2005chaplygin}. This Chaplygin gas (CG) model changes from a pressureless fluid
model to an accelerated expansion model as the scale factor varies from small to large values. Then comes its generalization, known as the generalized Chaplygin gas (GCG) \cite{alam2003exploring,gorini2003can,bento2002generalized,makler2003constraints}. After some modification, a new model known as the modified Chaplygin gas (MCG) model \cite{benaoum2012modified,debnath2004role} was created, which varies between the radiation stage (sometimes called the dust stage) and the $\Lambda CDM$ stage. Several other forms of CG have also evolved, such as variable Chaplygin gas \cite{guo2007cosmology}, viscous Chaplygin gas \cite{zhai2006viscous}, etc.\\

In 2003, a new version of CG was introduced, named the generalized cosmic Chaplygin gas (GCCG) \cite{gonzalez2003you}. It has an eccentric property in that it is stable and does not show any kind of unphysical behavior, even though the vacuum fluid is in the phantom era. In recent years, several investigations have been conducted on GCCG \cite{chakraborty2007generalized,rudra2013role,bhadra2012accretion,sharif2014effects,rudra2012dynamics,sharif2014phantom,eid2018schwarzschild,sharif2013reissner}. Debnath \cite{debnath2007variable} has introduced a new concept of variable modified Chaplygin gas (VMCG) which acts as a mixture in almost every stage of the universe and there is an intermediate stage where it becomes pressureless, and the content of matter becomes equivalent to pure dust. Numerous researches have been carried out on topics like thermodynamics of VMCG \cite{panigrahi2016thermodynamics}, its evolution on FRW universe \cite{chakraborty2019evolution}, the interaction of VMCG in a non-flat universe \cite{jamil2008interacting}, correspondence between VMCG and holographic dark energy \cite{chattopadhyay2009holographic} etc. The new variable-modified Chaplygin gas (NVMCG) is another variant of DE, which is an extension of MCG that is fascinating in terms of physical importance. Chakraborty and Debnath first introduced this concept of NVMCG \cite{chakraborty2010new}, Bhadra and Debnath have shown accretion procedure of NVMCG onto Schwarzschild and Kerr-Newman BHs \cite{bhadra2012accretion}, effectiveness of NVMCG as a DE has also been investigated \cite{rudra2013effective}. In 2021, Debnath first introduced two highly fascinating hypothetical variants of CG, named modified Chaplygin-Jacobi gas(MCJG) and modified Chaplygin-Abel gas(MCAG); both are appropriate for playing the role of DE in contributing to measuring the accelerated rate of expansion of our universe \cite{debnath2021roles}. Later, these two models have been discussed in the context of f(T) gravity inside a non-flat FLRW universe \cite{chaudhary2023constraints} and their accretion procedure onto Schwarzschild BH in FRW universe \cite{mukherjee2023accretion}. In addition, MCJG has been analyzed in the context of a non-commutative wormhole \cite{paramanik2024non}.\\

One of the most appealing studies related to BHs has always been its mass accretion procedure. Michel \cite{michel1972accretion} first established the equations of motion of spherically symmetric flow off and on condensed objects. Then Babichev et al. advanced the work by generalizing this concept more \cite{babichev2004black,babichev2005accretion}. The accretion process and condition of reducing into a naked singularity of charged BHs have been discussed \cite{jamil2008charged}; accretion phenomena of Kerr-Newman BH due to dark energy accretion has been studied in detail \cite{bhadra2012accretion,jimenez2008evolution}; dynamical behavior of phantom-like energy onto a stringy, magnetically charged BH has been derived  \cite{sharif2012phantom}; accretion, as well as evaporation process of modified Hayward BH, has been analyzed \cite{debnath2015accretion} and the process of phantom energy accretion of BHs in cyclic cosmology has also been studied \cite{sun2008phantom}.\\

Taking motivation from this kind of research, cosmologists started to explore higher-dimensional BHs and their characteristics. The phantom energy accretion process on a 3D BH in Einstein-Power-Maxwell gravity has been discussed \cite{abbas2013thermodynamics}; accretion procedure of higher-dimensional Schwarzschild BH has been analyzed \cite{john2013accretion}; (n+1)-dimensional charged BTZ-like BH solutions, their thermodynamics, stability, etc. have been discussed \cite{hendi2011charged}; entropy and scalar field structure of the D-dimensional Reissner-Nordstrom black hole have been studied \cite{kim1997renormalized}; also Debnath briefly investigated the accretion procedure of dark energy and dark matter onto (n+2)-dimensional BHs \cite{debnath2015accretions}.\\

Furthermore, one of the most discussed cosmology topics in recent times is Ho{\v{r}}ava-Lifshitz gravity. It is a nonrelativistic theory of gravity, which is power-counting, renormalizable, and a prospective way to recover Einstein's general relativity at large scales. This theory was first proposed by Ho{\v{r}}ava \cite{hovrava2009membranes,hovrava2009spectral}. Later, since even if it has an infrared (IR) fixed point, that is general relativity, it shows an anisotropic scaling of Lifshitz type in the ultraviolet (UV) between time and space, hence it is called Ho{\v{r}}ava-Lifshitz gravity. Its detailed discussion, extension \cite{cai2009dynamical} and also its application as the cosmological framework of the universe \cite{calcagni2009cosmology,kiritsis2009hovrava} have provided us with many possibilities. This theory is mainly popular as a postulant of the quantum field theory of gravity for the value of the degree of anisotropy between space and time, i.e. $z=3$. It provides us with the classical Einstein-Hilbert theory of gravity, in $(3+1)$-dimensions, with a fixed point in UV for $z=3$ and a fixed point in IR at $z=1$. So far, cosmologists have come up with only four possible types of HL gravity, namely, with/without the projectability condition and with/without the detailed balance condition, of which without detailed balance and with the projectability are the two most feasible conditions
\cite{bhattacharya2011thermodynamical}. HL gravity theory has many implications for solving problems regarding inflation and non-Gaussianity \cite{takahashi2009chiral,mukohyama2009scale}, creation of gravitational wave \cite{mukohyama2009phenomenological}, novel solution subclass
\cite{lu2009solutions,leon2009phase,minamitsuji2010classification}, dark energy phenomenology \cite{park2010test,setare2010holographic}, astrophysical phenomenology \cite{kim2009surplus,izumi2010stellar}, observational constraints \cite{dutta2009observational} etc. Considering all of this progress, researchers started to study black hole solutions in HL gravity \cite{lu2009solutions,kehagias2009black,cai2009topological,myung2009thermodynamics,park2009black}. Investigation on static, spherically symmetric black hole solutions in HL gravity has been done \cite{lu2009solutions}, Abbas \cite{abbas2014phantom} has shown the phantom energy accretion model onto a black hole in HL gravity. Also, logarithmic entropy of Kehagias-Sfetsos (KS) BH in IR modified Ho{\v{r}}ava gravity has been found \cite{liu2011logarithmic}, the first law of thermodynamics, as well as gravitational field equation for a static, spherically symmetric BH in with/without the detailed balance conditioned HL gravity, have been discussed in details \cite{cai2010horizon}, particle motion around a KS BH in deformed HL gravity has been investigated \cite{wei2019geodesics}, in addition entropy and mass for BHs with a general dynamical coupling constant `$\lambda$' by using the canonical Hamiltonian method in HL gravity have been analyzed \cite{cai2009thermodynamics}.\\

Recently, work related to constraining parameters in different kinds of gravity frameworks such as \cite{debnath2021roles,debnath2014constraining,debnath2021constraining,debnath2021observational,debnath2021gravitational,debnath2020gravitational,aberkane2017viscous,mukherjee2024constraining} has also gained considerable popularity among researchers.
Inspired by all of these amazing works mentioned above, here in this paper, we have discussed the mass accretion process of a 4-dimensional KS BH due to the presence of different types of DE candidates in the HL gravity scenario. We have constructed the paper in such a manner that in Sec.\ref{b}, we have obtained the equation of mass of a 4-dimensional KS BH in terms of the scale factor `$a$'. Then in Sec.\ref{c}, we have taken GCCG as a candidate for DE, and considering its energy density in terms of the dimensionless density parameter, we have established the relationship between the mass of BH `M' and the redshift function `z'. Similarly, in Sec.\ref{d}, we have taken VMCG; in Sec.\ref{e}, we have taken NVMCG; in Sec.\ref{f}, we have taken MCJG, and lastly in Sec.\ref{g}, we have taken MCAG as the candidate of DE and in each of the cases we have established the relationship between mass of the BH `M' and redshift function `z' by considering the energy density equation in terms of dimensionless density parameter. Finally, in Sect.\ref{h}, we have recapitulated all the findings of our work in a detailed manner.


\section{Accretion Phenomena of dark energy and dark matter onto Kehagias-Sfetsos (KS) Black Hole in Ho{\v{r}}ava-Lifshitz (HL) gravity:}\label{b}
Let us assume the scenario where our universe is governed by Ho{\v{r}}ava-Lifshitz (HL) gravity. Now, if we make use of the formulation by Arnowitt-Deser-Misner, in $(3+1)$-dimension, then the metric will be of the following form \cite{arnowitt1961coordinate}:

\begin{equation}\label{HL1}
 ds^2=-N^2dt^2+g_{_{lm}}(dx^l+N^ldt)(dx^m+N^mdt),
\end{equation}

Here, $N$ and $N_{l}$ are two dynamical variables known as the lapse rate and the shift function, respectively. Whereas $g_{_{lm}}$ is the spatial metric.\\
The gravitational action of the HL gravity is given as follows \cite{bhattacharya2011thermodynamical}:

\begin{equation}\label{HL2}
 S=\int{dtd^3x\sqrt{g}N(\mathcal{L_{KE}}+\mathcal{L_{VE}})}.
\end{equation}

where $\mathcal{L_{KE}}$ and $\mathcal{L_{VE}}$ denote the kinetic and potential energy parts of the Lagrangian, respectively.\\
Therefore, in the case of detailed balance-conditioned HL gravity, the above Eqn.(\ref{HL2}) will reduce into the following form\cite{kehagias2009black}:

\begin{multline}\label{HL3}
S=\int{dtd^3x\sqrt{g}N \left[\frac{2}{\kappa^2}\left(K_{lm}K^{lm}-\lambda
K^2\right)-\frac{\kappa^2}{2\omega^4}C_{lm}C^{lm}+\frac{\kappa^2\mu\epsilon^{lmt}}{2\omega^2}R_{lq}\bigtriangledown_{m}R^{q}_{t}\right.}\\
\left.-\frac{\kappa^2\mu^2}{8}R_{lm}R^{lm}+\frac{\kappa^2\mu^2}{8(1-3\lambda)}\left(\frac{1-4\lambda}{4}R^2+\Lambda
R-3\Lambda^2\right)+\mu^4R\right],     
\end{multline}

where, $\omega$, $\kappa$ and $\lambda$ symbolize coupling constants which are dimensionless. $\Lambda$ is the positive cosmological constant related to the IR limit and having mass dimension 2. In addition, $\epsilon^{lmt}$ stands for a totally antisymmetric tensor and $\mu$ is a parameter with mass dimension 1. For the fixed-time spacelike hypersurface, $K_{lm}$ is the extrinsic curvature, whose expression is given below:

\begin{equation}\label{HL4}
 K_{lm}=\frac{1}{2N}\left(\Dot{g}_{_{lm}}-\bigtriangledown_{l}N_{m}-\bigtriangledown_{m}N_{l}\right).
\end{equation}

Also, $C^{lm}$ is known as the Cotton tensor, given by the following form:

\begin{equation}\label{HL5}
C^{lm}=\epsilon^{ltq}\bigtriangledown_{t}\left(R^{m}_{q}-\frac{1}{4}R\delta^{m}_{q}\right).
\end{equation}

It should be noted that all of the covariant derivatives used here are defined by the spatial metric $g_{lm}$.
Now, if we use the projectability condition of HL gravity \cite{sotiriou2009quantum} and take the help of the Friedmann-Robertson-Walker (FRW) metric, that is:

\begin{equation}\label{HL6}
 N=1~,~g_{_{lm}}=a^2(t)\gamma_{_{lm}}~,~N^{l}=0~,
\end{equation}

and

\begin{equation}\label{HL7}
 \gamma_{_{lm}}dx^{l}dx^{m}=\frac{dr^2}{1-kr^2}+r^2d\Omega^{2}_{2}~.
\end{equation}\\

Where $k$ represents the curvature parameter, which can only take values $-1$, 0, or 1 according to the open, flat, or closed universe, respectively, and $a(t)$ is called the cosmic or Robertson-Walker scale factor. Let us suppose that the dark matter is in the form of a perfect fluid, then the Friedmann equation will transform into the following form \cite{kehagias2009black}: 

\begin{equation}\label{1}
 H^2=\frac{\kappa^2}{6(3\lambda-1)}\left[\rho-\frac{6k\mu^4}{a^2}-\frac{3k\kappa^2\mu^2}{8(3\lambda-1)a^4}\right],    
\end{equation}\\

Here, $H$ is the Hubble parameter, and $\rho$ is the usual energy density. [1mm]
In 2009, Kehagias and Sfetsos obtained an asymptotically flat and spherically symmetric spacetime with the same characteristics as the Schwarzschild BH at far distances from the source in Ho{\v{r}}ava gravity scenario \cite{kehagias2009black,culetu2015source}. This metric solution is known as Kehagias-Sfetsos(KS) Black Hole solutions in HL gravity with a singularity at $r=0$ whose divergence is quite milder than that of Schwarzschild BH's. Especially considering $\lambda=1$ \cite{kehagias2009black}, the above Eqn.(\ref{1}) will transform into the following form:

\begin{equation}\label{2}
 H^2=\frac{\kappa^2}{12}\left[\rho-\frac{6k\mu^4}{a^2}-\frac{3k\kappa^2\mu^2}{16a^4}\right].    
\end{equation}

It is quite clear from the above Eqn.(\ref{2}) that whenever $k=0$, the higher-order derivative terms will give zero contribution in action. In the case of $k\neq0$, they will have notable significance for small values of ``$a$" and will become less significant as the value of ``$a$" becomes large, which goes well with general relativity.\\[1mm]

Now, we are going to analyze the accretion phenomena of a 4-dimensional static KS BH; hence, let us consider the KS BH solution \cite{abbas2014phantom,liu2011logarithmic} given by:

\begin{equation}\label{3}
ds^2=\mathcal{A}(r)dt^2-\frac{\substack{1}}
{\substack{\mathcal{A}(r)}}dr^2-r^2(d\theta^2+sin^2\theta d\phi^2) ,
\end{equation}

where $\mathcal{A}(r)$ is given by:
\begin{equation*}\label{4}
  \mathcal{A}(r)=1+\omega r^2-\omega r^2\sqrt{1+\frac{4M}{\omega r^3}}~,
 \end{equation*}
 
and $   \omega=
 \frac{\substack{16\mu^2}}
 {\substack{\kappa^2}} $~
provided, here both $\mu$ and $\kappa$ are parameters related to HL gravity.\\

Invisaging in the form of a perfect fluid that follows the equation of state $p=p(\rho)$, the energy momentum tensor of the dark energy will be:

\begin{equation}\label{6}
  T_{rs}=(\rho+p)u_r u_s-pg_{rs} ,
 \end{equation}

Here, $\rho$ and $p$ represent the density of the dark energy and the pressure of the dark energy, respectively. The 4-velocity of the fluid $u^r=\frac{dx^r}{ds}$, satisfies the equation $u^r u_r=1$. Suppose that the falling dark energy has no influence on the spherical symmetry of the black hole.

Using the equation for energy conservation, which is $T^{0r}_{;r}=0$, we will have the following equation as \cite{debnath2015accretion}:

\begin{equation}\label{7}
 ur^2(\rho+p)\left(\sqrt{u^2+\mathcal{A}}\right)M^{-2}=D_0 ,
\end{equation}

where, $D_0$ symbolizes an integration constant.\\[1.5mm]

Also, $u_rT^{rs}_{;s}=0$, which is the equation of energy flux, obtained by the projection of the energy-momentum conservation law on the fluid's four-velocity, will give us the following equation \cite{debnath2015accretion}:

\begin{equation}\label{8}
 ur^2M^{-2}exp\left[\int_{\rho_\infty}^{\rho}{\frac{d\rho\prime}{\rho\prime+p(\rho\prime)}}\right]=-D_1 ,
\end{equation}

where $D_1>0$ symbolizes another integration constant related to the energy flux \cite{babichev2004black,babichev2013black,babichev2005accretion,debnath2015accretion}, has been evaluated in the Appendix in a detailed manner. Whereas $\rho$ and $\rho_\infty$ represent the values of the density of the dark energy at finite and infinite $r$.\\

Furthermore, the division of Eqn.(\ref{7}) by Eqn.(\ref{8}), leads us to the following equation \cite{debnath2015accretion}:

\begin{equation}\label{9}
 (\rho+p)\left(\sqrt{u^2+\mathcal{A}}\right)exp\left[-\int_{\rho_\infty}^{\rho}{\frac{d\rho\prime}{\rho\prime+p(\rho\prime)}}\right]=D_2 ,
\end{equation}

Here, $D_2=-\frac{D_0}{D_1}=[\rho_\infty+p(\rho_\infty)]$.\\[1mm]
Again, integration on the dark energy flux over the 4-dimensional black hole's volume will provide us the equation of the rate of change of mass as follows \cite{john2013accretion,debnath2015accretions}:

\begin{equation}\label{10}
\Dot{M}=-4\pi D_0 M^2 ,
\end{equation}

So, after substituting the value of $D_0$ in Eqn.(\ref{10}), we will get the equation of the rate of change of the mass of the black hole as follows \cite{babichev2011perfect}:

\begin{equation}\label{11}
\Dot{M}=4\pi D_1 M^2\left[\rho_\infty+p(\rho_\infty)\right] ,
\end{equation}

Following Refs.\cite{jamil2011accretion,jamil2011generalized,dutta2019dark,babichev2011perfect,abbas2014phantom} here we can easily conclude that the above Eqn.(\ref{11}) will give us the mass of the black hole for every $p$ and $\rho$, which does not satisfy the dominant energy condition, whenever we apply the equation of state $p=\omega\rho$.\\

Thus, the above Eqn.(\ref{11}) reduces to the following form:

\begin{equation}\label{12}
 \Dot{M}=4\pi D_1 M^2(\rho+p) .
\end{equation}

From the above Eqn.(\ref{12}), we can clearly observe that the rate of change of mass of the black hole exclusively depends upon the term $(\rho+p)$.In the case of quintessence kind of dark energies $(\rho+p)>0$, which implies $\Dot{M}>0$ that indicates the gradual rise of the mass of the black hole due to quintessence kind of dark energy's accretion onto it.Again, in the case of phantom kind of dark energies $(\rho+p)<0$, which implies $\Dot{M}<0$ that indicates the gradual decrease of the mass of the black hole due to phantom kind of dark energy's accretion onto it.\\[0.5mm]

Applying the conservation equation to a 4-dimensional black hole, we will get the following equation:

\begin{equation}\label{13}
 \Dot{\rho}+3H(\rho+p)=0.
\end{equation}



Now, considering that our universe combines dark energy and dark matter, its energy density and pressure will be the sum of energy densities and pressures of dark energy and dark matter, respectively, which can be written as follows:

\begin{equation*}\label{15}
\rho=\rho_m+\rho_D,
\end{equation*}

and

\begin{equation*}\label{16}
 p=p_m+p_D .
\end{equation*}\\

Here, $\rho_m$ and $\rho_D$ represent the energy density of dark matter and dark energy, respectively.
In a similar manner, $p_m$ and $p_D$ represent the pressure of dark matter and dark energy, respectively.\\

Following the above analogy and using the conservation of both dark matter and dark energy individually, also considering the redshift equation $1+z=\frac{1}{a}$ and the equation of state for dark matter $p_m=\omega_{_{ma}}\rho_m$, we will get the energy density of dark matter as follows \cite{debnath2015accretions}:

\begin{equation}\label{17}
 \rho_m=\rho_{_{m_0}}(1+z)^{3(1+\omega_{_{ma}})},
\end{equation}

Where $\rho_{_{m_0}}$ represents the current value of dark matter's energy density.\\

In terms of the dimensionless density parameter, $\Omega_{_{m0}}=\frac{\rho_{_{m0}}}{3H^2_0}$~, the above equation will transform into the following form:

\begin{equation}\label{density_dark matter}
 \rho_m=3H^2_0\Omega_{_{m0}}(1+z)^{3(1+\omega_{_{ma}})}.   
\end{equation}

Now, using Eqn.(\ref{13}) into Eqn.(\ref{12}) and taking integration on both sides, we will obtain the general equation of mass of a 4-dimensional black hole as follows:

\begin{equation}\label{mass_BH}
M=\frac{M_0}{1+\frac{4\pi D_1 M_0}{3}\mathop{\mathlarger{\int}_{\rho_0}^{\rho}}{\frac{d\rho}{H}}},
\end{equation}

Here, $M_0$ stands for the present mass of the black hole, and $\rho_0$ is the present energy density is given by $\rho_0=\rho_{m0}+\rho_{D0}$ \cite{debnath2015accretions,geng2015accretion}, where $\rho_{m0}$ and $\rho_{D0}$ are the present values of the dark matter and dark energy densities respectively.\\[1mm]
Finally, using Eqn.(\ref{2}) into the above Eqn.(\ref{mass_BH}), we will get our required equation of mass for a 4-dimensional Kehagias-Sfetsos(KS) black hole in Ho{\v{r}}ava-Lifshitz (HL) gravity scenario as follows:

\begin{equation}\label{18}
M=\frac{M_0}{1+\frac{4\pi D_1 M_0}{3}\mathop{\mathlarger{\int_{\rho_0}^{\rho}}}\frac{d\rho}{\sqrt{\frac{\kappa^2}{12}\left[\rho-\frac{6k\mu^4}{a^2}-\frac{3k\kappa^2\mu^2}{16a^4}\right]}}}.    
\end{equation}

So, Eqn.(\ref{18}) is the equation of mass that we have desired to obtain for a 4-dimensional Kehagias-Sfetsos(KS) BH in Ho{\v{r}}ava-Lifshitz (HL) gravity. The corresponding constant $D_1$ has been obtained by Eqn.(\ref{ICE11}) in the Appendix.


\subsection{Generalized Cosmic Chaplygin Gas (GCCG)}\label{c}
The Chaplygin gas (CG) model as a candidate for dark energy to justify the universe's accelerated expansion is very popular among cosmologists. So, with time, a considerable number of investigations have been carried out on the Chaplygin gas \cite{kamenshchik2001alternative}, and it has inspired cosmologists to explore modified and generalized versions of this hypothetical gas, such as the generalized Chaplygin gas (GCG)
\cite{gorini2003can} and modified Chaplygin gas (MCG) \cite{benaoum2012modified,debnath2004role}. This kind of generalization can have quite a remarkable effect on the early stage of the universe. If we consider the generalizations of the cosmic Chaplygin gas (CCG) model together with an initial parameter $w$, which is adjustable in such a way that even if the vacuum fluid obeys the condition of phantom energy, the subsequent models are free from unreal behaviors and can have stability. It is an example of how we can avoid the Big-Rip singularity and the evaporation of the black hole mass to zero. It has the following equation of state \cite{gonzalez2003you,chakraborty2007generalized}:

\begin{equation}\label{19}
 p=-\rho^{-\alpha}\left[G+(\rho^{1+\alpha}-G)^{-w}\right] ,
\end{equation}

where

\begin{equation*}\label{20}
    G=\frac{E}{1+w}-1 .
\end{equation*}\\[1mm]

Here, $E$ is a constant that has both positive and negative values, and $-q~<~w~<~0$~, $q$ is a positive definite constant that can take any value greater than unity. Also, $\alpha$ is a constant which belongs to [0,1].\\[1.5mm]
The equation of state for this model satisfies the following four conditions together with the fact that whenever $w\to0$, it will reduce to unified models of dark energy and dark matter of current Chaplygin Gas:
(i) When $w=-1$ or at a late time, the equation of state reduces to a de Sitter fluid, (ii) Whenever the Chaplygin parameter $E\to0$, the equation of state becomes $p=w\rho$, (iii) When energy density is high, it converts into the equation of state for current Chaplygin unified dark matter models and lastly (iv) At the late time, because the involved parameters have physically reasonable values, the development for density perturbations obtained from the selected equation of state does not affect the pathological behavior of the matter power spectrum.The equation of state given by Eqn.(\ref{19}) indicates the dust era in the past and $\Lambda CDM$ era in the future. This model is widely known as the Generalized Cosmic Chaplygin Gas (GCCG) model.\\[1mm]
The major benefit of the GCCG model is that it is more flexible than the GCG and MCG models. In addition, it can be applied to any realm of
cosmology simply by changing the initial parameter $w$ suitably.\\
The energy density for this model is given by \cite{gonzalez2003you,chakraborty2007generalized}:

\begin{equation}\label{21}
 \rho=\left[G+\left(1+\frac{F}{a^{3(1+\alpha)(1+w)}}\right)^{\frac{1}{1+w}}\right]^{\frac{1}{1+\alpha}} ,
\end{equation}

Here $F$ is an integrating constant that can only take positive values. [0.5mm] Now, the energy density required for GCCG accretion will be given by:

\begin{equation}\label{22}
\rho_{_{GCCG}}=\rho_{_{m_0}}(1+z)^{3(1+\omega_{_{ma}})}+\left[G+\left(1+F(1+z)^{3(1+\alpha)(1+w)}\right)^{\frac{1}{1+w}}\right]^{\frac{1}{1+\alpha}},
\end{equation}\\[0.5mm]

 In terms of dimensionless density parameter, $\Omega_{_{GCCG0}}=\frac{\rho_{_{GCCG0}}}{3H^2_0}$ and using Eqn.(\ref{density_dark matter}) we can express the above equation as follows \cite{debnath2021observational}:

\begin{equation}\label{density_GCCG}
\rho_{_{GCCG}}=3H^2_0\left[\Omega_{_{m0}}(1+z)^{3(1+\omega_{_{ma}})}+\Omega_{_{GCCG0}}\left\{\widetilde{A} +(1-\widetilde{A})
\left(\widetilde{A}_{s}+(1-\widetilde{A}_{s})(1+z)^{3(1+\alpha)(1+w)}\right)^{\frac{1}{1+w}}\right\}^{\frac{1}{1+\alpha}}\right]. 
\end{equation}

where, $\rho_{_{GCCG0}}$ gives us the present energy density of GCCG. Again, the other parameters have the following expressions: $\widetilde{A}_{s}=\frac{1}{1+F}$,
$\widetilde{A}=\left(1+G^{-1}\widetilde{A}_{s}^{-\frac{1}{1+w}}\right)^{-1}$,
and $\rho_{_{GCCG0}}^{1+\alpha}=G+\widetilde{A}_{s}^{-\frac{1}{1+w}}$.\\

So, putting the above Eqn.(\ref{density_GCCG}) into the Eqn.(\ref{18}), we will get the equation of mass in terms of the redshift function $z$ as follows:

\fontsize{7pt}{9pt}\selectfont
\begin{equation}\label{23}
  M=\frac{M_0}{1+\frac{4\pi D_1 M_0}{3}\mathop{\mathlarger{\int}_{\rho_{_{GCCG0}}}^{\rho_{_{GCCG}}}}\frac{d\left(3H^2_0\left[\Omega_{_{m0}}(1+z)^{3(1+\omega_{_{ma}})}+\Omega_{_{GCCG0}}\left\{\widetilde{A} +(1-\widetilde{A})
\left(\widetilde{A}_{s}+(1-\widetilde{A}_{s})(1+z)^{3(1+\alpha)(1+w)}\right)^{\frac{1}{1+w}}\right\}^{\frac{1}{1+\alpha}}\right]\right)}{\sqrt{\frac{\kappa^2}{12}\left[\left(3H^2_0\left[\Omega_{_{m0}}(1+z)^{3(1+\omega_{_{ma}})}+\Omega_{_{GCCG0}}\left\{\widetilde{A} +(1-\widetilde{A})
\left(\widetilde{A}_{s}+(1-\widetilde{A}_{s})(1+z)^{3(1+\alpha)(1+w)}\right)^{\frac{1}{1+w}}\right\}^{\frac{1}{1+\alpha}}\right]\right)-6k\mu^4(1+z)^2-\frac{3k\kappa^2\mu^2(1+z)^4}{16}\right]}}}.
\end{equation}\\
\fontsize{10pt}{12pt}\selectfont

\begin{figure}[H]
\begin{subfigure}{.4\textwidth}
    \includegraphics[height=2.0in]{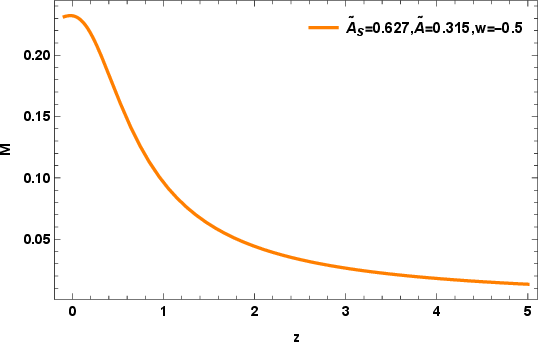}
    \subcaption{For the values: $\bf{\widetilde{A}_{s}=0.627}$,$\bf{\widetilde{A}=0.315}$ and $\bf{w=-0.5}$}
     \label{fig:f 1}
      \end{subfigure}
      \hfill
      \begin{subfigure}{.4\textwidth}
    \includegraphics[height=2.0in]{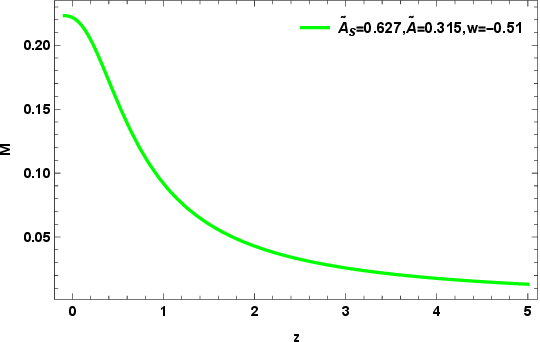}
    \subcaption{For the values: $\bf{\widetilde{A}_{s}=0.627}$,$\bf{\widetilde{A}=0.315}$ and $\bf{w=-0.51}$}
     \label{fig:f 1}
      \end{subfigure}
  \begin{subfigure}{.4\textwidth}
    \includegraphics[height=2.0in]{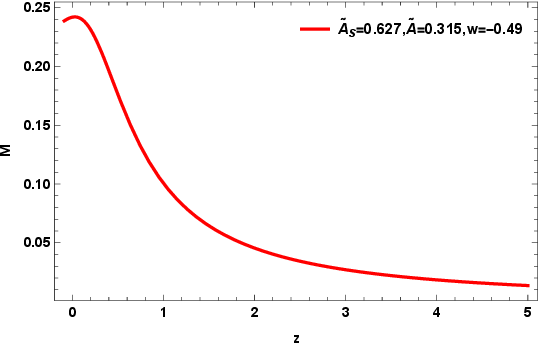}
    \subcaption{For the values: $\bf{\widetilde{A}_{s}=0.627}$,$\bf{\widetilde{A}=0.315}$ and $\bf{w=-0.49}$}
     \label{fig:f 1}
      \end{subfigure}
      \hfill
      \begin{subfigure}{.4\textwidth}
    \includegraphics[height=2.0in]{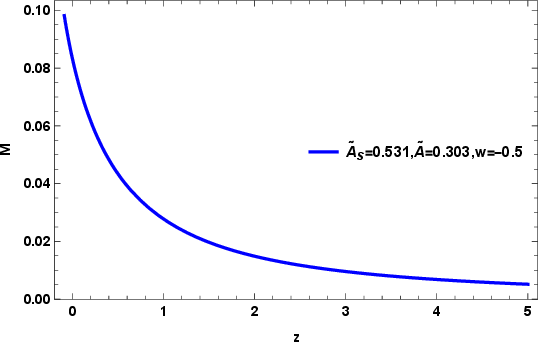}
    \subcaption{For the values: $\bf{\widetilde{A}_{s}=0.531}$,$\bf{\widetilde{A}=0.303}$ and $\bf{w=-0.5}$}
     \label{fig:f 1}
      \end{subfigure}
    \vfill
     \begin{subfigure}{.4\textwidth}
    \includegraphics[height=2.0in]{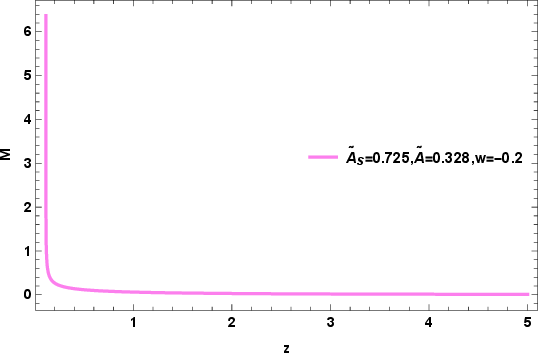}
    \subcaption{For the values: $\bf{\widetilde{A}_{s}=0.725}$,$\bf{\widetilde{A}=0.328}$ and $\bf{w=-0.2}$}
     \label{fig:f 1}
      \end{subfigure}
    \hfill
    \begin{subfigure}[b]{.4\textwidth}
    \includegraphics[height=2.0in]{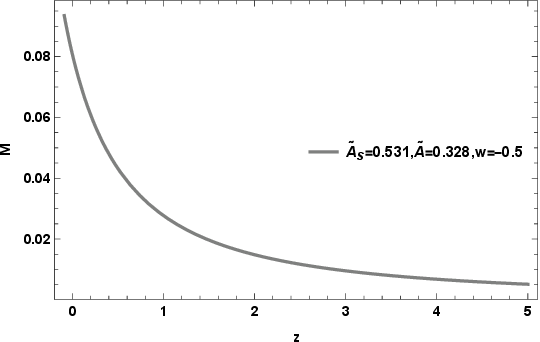}
    \subcaption{For the values: $\bf{\widetilde{A}_{s}=0.531}$,$\bf{\widetilde{A}=0.328}$ and $\bf{w=-0.5}$}
     \label{fig:f 1}
      \end{subfigure}
\caption{The variations in BH mass $M$ against redshift $z$ due to the accretion of GCCG for different values of parameters $\bf{\widetilde{A}_{s}}$, $\bf{\widetilde{A}}$ and $\bf{w}$.[Other parameters: $\kappa=1$~,~$\mu=1$~,~$k=1$~,~$\omega_{_{ma}}=0.01$~,~$\Omega_{_{mo}}=0.21$~,~$\Omega_{_{GCCG0}}=0.68$~,~$H_0=72$~,~$M_0=10$~,~$\rho_\infty=0.1$~and~$\alpha=0.060$]}
    \label{fig:my_label1}
      \end{figure}

The above Eqn.(\ref{23}) is the desired mass equation of a 4-dimensional KS BH in the HL gravity scenario due to the accretion of GCCG. The corresponding constant $D_1$ has been obtained by Eqn.(\ref{ICE15}) in the Appendix.\\[1.5mm]  
The depiction of the BH mass $M$ against the redshift $z$ has been shown through Fig.\ref{fig:my_label1} for different values of parameters $\widetilde{A}_{s}$, $\widetilde{A}$ and $w$. Hence it is quite evident from Fig.\ref{fig:my_label1} that, due to the accretion of GCCG, there is an enhancement in the mass of a 4-dimensional KS BH, i.e., with the successive rise of the universe, the mass of BH will also increase gradually because of the GCCG accretion in the HL gravity scenario.

\subsection{Variable Modified Chaplygin Gas (VMCG)}\label{d}
During the evolution of research on the Chaplygin gas, Zhang and Guo came up with a new kind of Chaplygin gas (CG) model named the variable Chaplygin gas (VCG) model \cite{guo2005observational,guo2007cosmology}, where the constant term in the equation of state of the Chaplygin gas has changed into a variable of cosmological scale factor `$a$'. This modification works fine in both the current accelerated expansion phase as well as the dust phase of the universe. However, it cannot justify the prior radiation stage of our universe. So, cosmologists introduced another model, named the modified Chaplygin gas (MCG) model \cite{benaoum2012modified,debnath2004role}.\\
The equation of state for MCG is given as follows:

\begin{equation}\label{24}
   p=P\rho-\frac{K}{\rho^\alpha} ,
\end{equation}

where, $P$ and $K$ are constants with positive values and $\alpha\in[0,1]$.
In this MCG model, all three major stages of our universe, namely radiation, dust, and the ongoing accelerated expansion phase, mainly influenced by dark energy, have been explained in a very good manner. Later, to obtain greater accuracy in the observational data, Debnath first proposed a model named the
variable modified Chaplygin gas (VMCG) model by replacing the constant $K$, in the above Eqn.(\ref{24}) with a positive function of the cosmological scale factor `$a$' as given by \cite{debnath2007variable}:

\begin{equation}\label{25}
p=P\rho-\frac{K(a)}{\rho^\alpha} ,
\end{equation}

with $P$ being a constant $>0$, $K(a)$ is some positive function of the cosmological scale factor `$a$' and $\alpha\in[0,1]$.\\[0.5mm]
For simplification, if we consider ;

\begin{equation}\label{26}
K(a)=Qa^{-m} .
\end{equation}\\

where, $Q>0$ and m are constants.\\[0.5mm]
Again, for this model, the equation of energy density is given as follows \cite{debnath2007variable}:

\begin{equation}\label{27}
\rho=\left[\frac{3(1+\alpha)Q}{(3(1+\alpha)(1+P)-m)}\frac{1}{a^m}+\frac{R}{a^{3(1+P)(1+\alpha)}}\right]^\frac{1}{1+\alpha} ,
\end{equation}\\

Here, $R$ is a positive integration constant. Also, the term $3(1+P)(1+\alpha)$  must be greater than $m$; otherwise, we cannot ensure the positivity of the first term. Again, $m$ should be positive else; it would lead to $a\to\infty$ giving $\rho\to\infty$, contradicting our theory of expanding the universe.\\[0.5mm]
Now, the required energy density for the accretion of VMCG will be given by:

\begin{equation}\label{28}
\rho_{_{VMCG}}=\rho_{_{m_0}}(1+z)^{3(1+\omega_{_{ma}})}+\left[\frac{3(1+\alpha)Q}{(3(1+\alpha)(1+P)-m)}(1+z)^m+R(1+z)^{3(1+P)(1+\alpha)}\right]^\frac{1}{1+\alpha} ,
\end{equation}\\

 Furthermore, in terms of dimensionless density parameter,  $\Omega_{_{VMCG0}}=\frac{\rho_{_{VMCG0}}}{3H^2_0}$ and using Eqn.(\ref{density_dark matter}) we can express the above equation as follows \cite{debnath2020gravitational,ranjit2016observational}:

\begin{equation}\label{density_VMCG}
\rho_{_{VMCG}}=3H^2_0\left[\Omega_{_{m0}}(1+z)^{3(1+\omega_{_{ma}})}+\Omega_{_{VMCG0}}\left\{{\widetilde{B}_{s}(1+z)^m+(1-\widetilde{B}_{s})(1+z)^{3(1+P)(1+\alpha)}}\right\}^\frac{1}{1+\alpha}\right].   
\end{equation}\\

Here, $\rho_{_{VMCG0}}$ stands for the current energy density of the VMCG. Also, $\widetilde{B}_{s}=1-\frac{R}{\rho_{_{VMCG0}}^{1+\alpha}}$ and $\rho_{_{VMCG0}}^{1+\alpha}=R+\frac{3(1+\alpha)Q}{3(1+\alpha)(1+P)-m}$.\\[0.2mm]

So, putting the above Eqn.(\ref{density_VMCG}) into the Eqn.(\ref{18}), we will get the equation of mass in terms of the redshift function $z$ as
follows:

\fontsize{9pt}{11pt}\selectfont
\begin{equation}\label{29}
M=\frac{M_0}{1+\frac{4\pi D_1 M_0}{3}\mathop{\mathlarger{\int}}_{\rho_{_{VMCG0}}}^{\rho_{_{VMCG}}}\frac{d\left(3H^2_0\left[\Omega_{_{m0}}(1+z)^{3(1+\omega_{_{ma}})}+\Omega_{_{VMCG0}}\left\{{\widetilde{B}_{s}(1+z)^m+(1-\widetilde{B}_{s})(1+z)^{3(1+P)(1+\alpha)}}\right\}^\frac{1}{1+\alpha}\right]\right)}{\sqrt{\frac{\kappa^2}{12}\left[\left(3H^2_0\left[\Omega_{_{m0}}(1+z)^{3(1+\omega_{_{ma}})}+\Omega_{_{VMCG0}}\left\{{\widetilde{B}_{s}(1+z)^m+(1-\widetilde{B}_{s})(1+z)^{3(1+P)(1+\alpha)}}\right\}^\frac{1}{1+\alpha}\right]\right)-6k\mu^4(1+z)^2-\frac{3k\kappa^2\mu^2(1+z)^4}{16}\right]}}}.    
\end{equation}
\fontsize{10pt}{12pt}\selectfont

The above Eqn.(\ref{29}) is the desired mass equation of a 4-dimensional KS BH in the HL gravity scenario due to the accretion of VMCG. The corresponding constant $D_1$ has been obtained by Eqn.(\ref{ICE19}) in the Appendix.

\begin{figure}[H]
    \centering
    \includegraphics[width=0.45\linewidth]{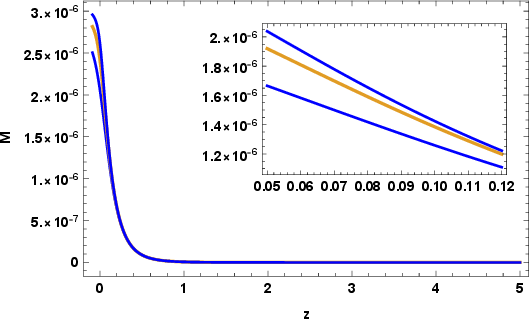}
    \caption{The variations in BH mass $M$ against redshift $z$ due to the accretion of VMCG for different values of  $\bf{\alpha = 0~,~0.5~,~1}$ corresponding to the down-to-up order of the curves. [Other parameters: $\kappa=1$~,~$\mu=1$~,~$k=1$~,~$\omega_{_{ma}}=0.01$~,~$\Omega_{_{mo}}=0.21$~,~$\Omega_{_{VMCG0}}=0.78$~,~$H_0=72$~,~$M_0=10$~,~$m=0.5$~,~$P=5$~,~$Q=10~$,~$\rho_\infty=0.1$~and~$\widetilde{B}_{s}=0.0532$]}
\label{fig:my_label2}
\end{figure}

\begin{figure}[H]
 \begin{subfigure}{.4\textwidth}
    \includegraphics[height=2.0in]{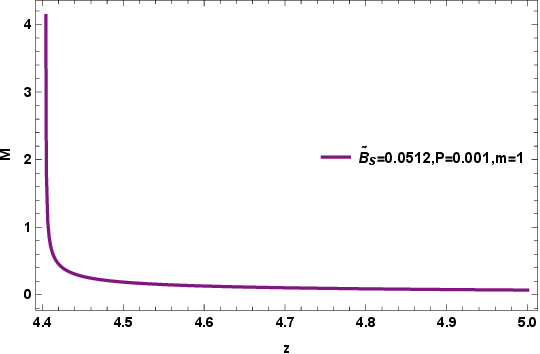}
    \subcaption{For the values: $\bf{\widetilde{B}_{s}=0.0512}$, $\bf{P=0.001}$ and $\bf{m=1}$}
     \label{fig:f 3}
      \end{subfigure}
      \hfill
      \begin{subfigure}{.4\textwidth}
    \includegraphics[height=2.0in]{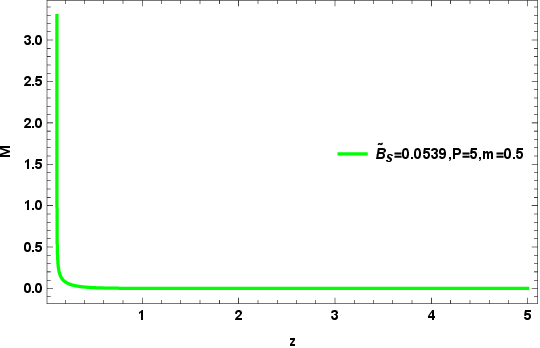}
    \subcaption{For the values: $\bf{\widetilde{B}_{s}=0.0539}$, $\bf{P=5}$ and $\bf{m=0.5}$}
     \label{fig:f 3}
      \end{subfigure}
    \vfill
    \center
     \begin{subfigure}{.4\textwidth}
    \includegraphics[height=2.0in]{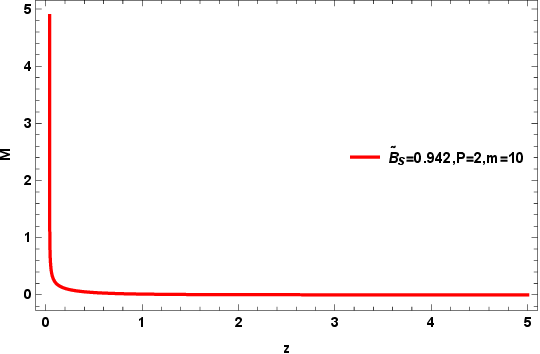}
    \subcaption{For the values: $\bf{\widetilde{B}_{s}=0.942}$, $\bf{P=2}$ and $\bf{m=10}$}
     \label{fig:f 3}
      \end{subfigure}
      \caption{The variations in BH mass $M$ against redshift $z$ due to the accretion of VMCG for different values of  parameter $\bf{\widetilde{B}_{s}}$ and constants~$\bf{P}$~,~$\bf{m}$.~[Other parameters: $\kappa=1$~,~$\mu=1$~,~$k=1$~,~$\omega_{_{ma}}=0.01$~,~$\Omega_{_{mo}}=0.21$~,~$\Omega_{_{VMCG0}}=0.78$~,~$H_0=72$~,~$M_0=10$~,~$Q=10$~,~$\rho_\infty=0.1$~and~$\alpha=0.5$]}
    \label{fig:my_label3}
\end{figure}

The depiction of the black hole mass $M$ against the redshift $z$ has been shown through Fig.\ref{fig:my_label2} for different values of $\alpha = 0~,~0.5~,~1$~corresponding to the down-to-up order of the curves. Also, in Fig.\ref{fig:my_label3}, the same has been plotted for different values of parameter $\widetilde{B}_{s}$ and constants $P$ and $m$. From Fig.\ref{fig:my_label2} and Fig.\ref{fig:my_label3}, it is quite evident that due to the accretion of VMCG, there is an enhancement in the mass of a 4-dimensional KS BH, i.e.,with the successive rise of the universe, the black hole mass will also increase gradually because of VMCG accretion in the HL gravity scenario.

\subsection{New Variable Modified Chaplygin Gas (NVMCG)}\label{e}
After the VMCG model \cite{debnath2007variable,chakraborty2019evolution}, researchers wanted to obtain a more heterogeneous model to explain the gradual development of our universe. In this regard, if we replace the constants in the equation of state of the MCG model, given by Eqn.(\ref{24}) with some positive functions of cosmological scale factor `$a$', then we will get the equation of state for the new variable modified Chaplygin gas (NVMCG) model as follows \cite{chakraborty2010new}:

\begin{equation}\label{30}
  p=S(a)\rho-\frac{T(a)}{\rho^\alpha},
\end{equation}

with $\alpha\in[0,1]$ and $S(a)$, $T(a)$ being some positive functions of the cosmological scale factor `$a$'.\\[0.5mm]
In particular, if we take $S(a)$ and $T(a)$ to be in the form:

\begin{equation}\label{31}
    S(a)=Pa^{-j},
\end{equation}

and

\begin{equation}\label{32}
    T(a)=Qa^{-i}.
\end{equation}\\

where, $P$, $Q$, $i$ and $j$ are all positive constants.\\
It is quite obvious that whenever $i=j=0$, Eqn.(\ref{30}) will reduce to the equation of state for the MCG model \cite{debnath2004role}. Also, if only $j=0$, then Eqn.(\ref{30}) will reduce to the equation of state for the VMCG model \cite{debnath2007variable}. The above model has some very compelling physical significance and can justify our universe's evolution process by adequately choosing parametric values.\\
Now, the equation of energy density for this model is given by \cite{chakraborty2010new}:

\begin{equation}\label{33}
  \rho=a^{-3}e^{\frac{3Pa^{-j}}{j}}\left[R+\frac{Q}{P}\left(\frac{3P(1+\alpha)}{j}\right)^\frac{3(1+\alpha)+j-i}{j}\Gamma\left({\frac{i-3(1+\alpha)}{j},\frac{3P(1+\alpha)}{j}a^{-j}}\right)\right]^\frac{1}{1+\alpha},
\end{equation}\\

Here, $\Gamma(b,z)$ represents the upper incomplete gamma function, while $R$ stands for an integration constant.\\[1mm]
Now, the energy density required for the accretion of NVMCG will be given by:

\begin{equation}\label{34}
 \rho_{_{NVMCG}}=\rho_{_{m_0}}(1+z)^{3(1+\omega_{_{ma}})}+(1+z)^3e^{\frac{3P(1+z)^j}{j}}\left[R+\frac{Q}{P}\left(\frac{3P(1+\alpha)}{j}\right)^\frac{3(1+\alpha)+j-i}{j}\Gamma\left({\frac{i-3(1+\alpha)}{j},\frac{3P(1+\alpha)}{j}(1+z)^j}\right)\right]^\frac{1}{1+\alpha}, 
\end{equation}\\

Now, making use of the Eqn.(\ref{density_dark matter}) and converting the above Eqn.(\ref{34}) in terms of dimensionless density parameter,  $\Omega_{_{NVMCG0}}=\frac{\rho_{_{NVMCG0}}}{3H^2_0}$~, we will get the following form of the energy density equation as:

\begin{multline}\label{density_NVMCG}
 \rho_{_{NVMCG}}=3H^2_0\left[\Omega_{_{m0}}(1+z)^{3(1+\omega_{_{ma}})}
 +\Omega_{_{NVMCG0}}(1+z)^3e^{\frac{3P[(1+z)^j-1]}{j}}\right.\\
\left. \left\{1-\widetilde{C}_{s}~\Gamma\left(\frac{i-3(1+\alpha)}{j},\frac{3
P(1+\alpha)}{j}\right)+\widetilde{C}_{s}~\Gamma\left(\frac{i-3(1+\alpha)}{j},\frac{3
P(1+\alpha)}{j}~(1+z)^{j}\right)\right\}^{\frac{1}{1+\alpha}}\right].   
\end{multline}\\[0.2mm]

Here, $\rho_{_{NVMCG0}}$ stands for the current energy density of NVMCG. The parameters used in the above equation has the following forms:~$\widetilde{C}_{s}=\frac{\widetilde{C}}{R+\widetilde{C}\widetilde{D}}$~, $\rho_{_{NVMCG0}}=e^{\frac{3P}{j}}\left[R+\widetilde{C}\widetilde{D}\right]^{\frac{1}{1+\alpha}}$~, $\widetilde{C}=\frac{Q}{R}\left(\frac{3P(1+\alpha)}{j}\right)^{\frac{(3(1+\alpha)+j-i)}{j}}$~~and
$\widetilde{D}=\Gamma\left(\frac{i-3(1+\alpha)}{j},\frac{3P(1+\alpha)}{j}\right)$.\\

Hence, from the Eqn.(\ref{18}), we will get the equation of mass in terms of the redshift function $z$ as follows:

\begin{equation}\label{35}
M=\frac{M_0}{1+\frac{4\pi D_1 M_0}{3}\mathop{\mathlarger{\int}}_{\rho_{_{NVMCG0}}}^{\rho_{_{NVMCG}}}\frac{d\rho_{_{NVMCG}}}{\sqrt{\frac{\kappa^2}{12}\left[\rho_{_{NVMCG}}-6k\mu^4(1+z)^2-\frac{3k\kappa^2\mu^2(1+z)^4}{16}\right]}}}.    
\end{equation}\\

where, the expression for $\rho_{_{NVMCG}}$ is explicitly given by the Eqn.(\ref{density_NVMCG}).\\

The above Eqn.(\ref{35}) is the desired mass equation of a 4-dimensional KS BH in the HL gravity scenario due to the accretion of NVMCG. The corresponding constant $D_1$ has been obtained by Eqn.(\ref{ICE23}) in the Appendix.

\begin{figure}[H]
    \centering
    \includegraphics[width=0.45\linewidth]{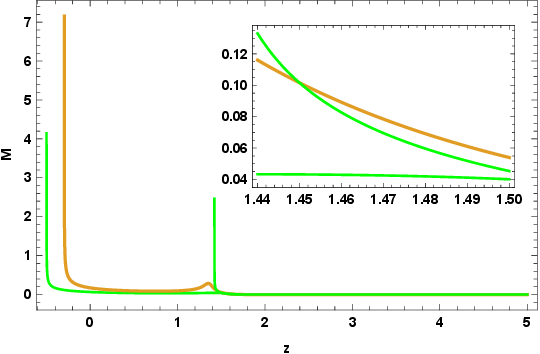}
    \caption{The variations in BH mass $M$ against redshift $z$ due to the accretion of NVMCG for different values of $\bf{\alpha = 0.1,0.5,0.9}$ corresponding to down to up order of curves. [Other parameters: $\kappa=1$~,~$\mu=1$~,~$k=1$~,~$\omega_{_{ma}}=0.01$~,~$\Omega_{_{mo}}=0.21$~,~$\Omega_{_{NVMCG0}}=0.75$~,~$H_0=72$~,~$M_0=10$~,~$\rho_\infty=0.1$~,~$\widetilde{C}_{s}=0.62$~,~$P=0.0000868$~,~$Q=2.988$~,~$i=7$~and~$j=13$]}
\label{fig:my_label4}
\end{figure}

\begin{figure}[H]
 \begin{subfigure}{.4\textwidth}
    \includegraphics[height=2.0in]{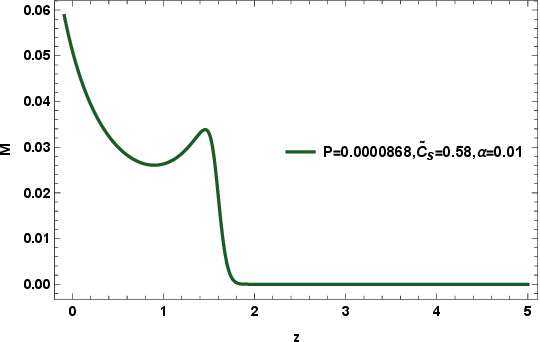}
    \subcaption{For the values: $P=0.0000868$, $\widetilde{C}_{s}=0.58$ and $\alpha=0.01$}
     \label{fig:f 5}
      \end{subfigure}
      \hfill
      \begin{subfigure}{.4\textwidth}
    \includegraphics[height=2.0in]{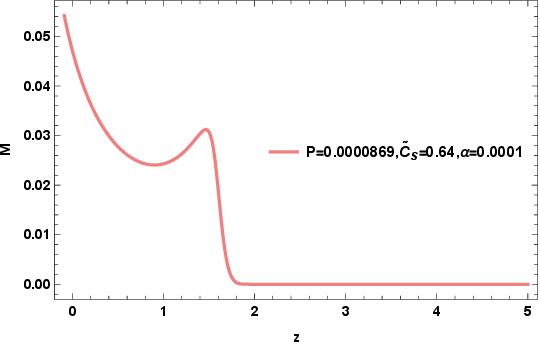}
    \subcaption{For the values:  $P=0.0000869$, $\widetilde{C}_{s}=0.64$ and $\alpha=0.0001$}
     \label{fig:f 5}
      \end{subfigure}
      \caption{The variations in BH mass $M$ against redshift $z$ due to the accretion of NVMCG for different values of  parameter~$\bf{\widetilde{C}_{s}}$~and constants~$\bf{P}$~and~$\bf{\alpha}$.[Other parameters: $\kappa=1$~,~$\mu=1$~,~$k=1$~,~$\omega_{_{ma}}=0.01$~,~$\Omega_{_{mo}}=0.21$~,~$\Omega_{_{NVMCG0}}=0.75$~,~$H_0=72$~,~$M_0=10$~,~$Q=2.988$~,~$\rho_\infty=0.1$~,~$i=7$~and~$j=13$]}
    \label{fig:my_label5}
\end{figure}

The depiction of the BH mass $M$ against the redshift $z$ has been shown through Fig.\ref{fig:my_label4} for different values of $\alpha = 0.1~,~0.5~,~0.9$ corresponding to the down to up order of the curves. Also, in Fig.\ref{fig:my_label5}, the same has been plotted for different values of parameter~$\widetilde{C}_{s}$~and constants~$P$~and~$\alpha$~. From both Fig.\ref{fig:my_label4} and Fig.\ref{fig:my_label5}, it is quite evident that due to the accretion of NVMCG, there is an enhancement in the mass of a 4-dimensional KS BH, i.e., with the successive rise of the universe, black hole mass will also increase gradually because of NVMCG accretion in the HL gravity scenario.

\subsection{Modified Chaplygin-Jacobi Gas (MCJG)}\label{f}
Following the trend of generalization and modification of the Chaplygin gas as a potent candidate of dark energy, Villanueva has deduced a generalized Chaplygin-Jacobi gas (GCJG) model \cite{villanueva2015generalized} in 2015. In the formulation of the GCJG model, a particular form of elliptic function, called Jacobi's elliptic function \cite{villanueva2015jacobian}, has been used together with the generalized Chaplygin scalar field related to the Hubble parameter. After that, Debnath has deduced a model of modified Chaplygin-Jacobi gas (MCJG) by replacing the hyperbolic function present in the generating function of MCG model by Jacobi elliptic cosine function \cite{debnath2021roles}.\\
For the MCJG model, the relation of pressure with energy density is given as follows \cite{debnath2021roles}:

\begin{equation}\label{36}
 p=\left[(2l-1)(1+R)-1\right]\rho-\frac{lS}{\rho^\alpha}+\frac{(1-l)(1+R)^2}{S}\rho^{2+\alpha} ,
\end{equation}

Where, both $R$ and $S$ are positive constants, $l\in[0,1]$ is called the elliptic modulus and $\alpha\in(0,1)$ is another constant. From the above Eqn.(\ref{36}), it is easy to deduce that whenever $R=0$ , it will reduce to the equation of state for GCJG model\cite{villanueva2015generalized}.\\
Now, the expression for the energy density will be given by \cite{debnath2021roles}:

\begin{equation}\label{37}
 \rho=\left(\frac{S}{1+R}\right)^\frac{1}{1+\alpha}\left[\frac{a^{3(1+\alpha)(1+R)}+lP}{a^{3(1+\alpha)(1+R)}-(1-l)P}\right]^\frac{1}{1+\alpha} ,
\end{equation}

with $P$ being a positive constant.\\
Again, the required energy density for accretion of MCJG will be given by:

\begin{equation}\label{38}
 \rho_{_{MCJG}}=\rho_{_{m_0}}(1+z)^{3(1+\omega_{_{ma}})}+\left(\frac{S}{1+R}\right)^\frac{1}{1+\alpha}\left[\frac{(1+z)^{-3(1+\alpha)(1+R)}+lP}{(1+z)^{-3(1+\alpha)(1+R)}-(1-l)P}\right]^\frac{1}{1+\alpha} ,
\end{equation}\\

 Furthermore, if we apply Eqn.(\ref{density_dark matter}) and use dimensionless density parameter $\Omega_{_{MCJG0}}=\frac{\rho_{_{MCJG0}}}{3H^2_0}$~, then the above equation will be reduced to the following form \cite{debnath2021roles}:

\begin{equation}\label{density_MCJG}
\rho_{_{MCJG}}=3H^2_0\left[\Omega_{_{m0}}(1+z)^{3(1+\omega_{_{ma}})}
 +\Omega_{_{MCJG0}}\left[l\widetilde{D}_{s}-(1-l)(1-\widetilde{D}_{s})\right]^{\frac{1}{1+\alpha}}\left[\frac{\widetilde{D}_{s}+(1-\widetilde{D}_{s})(1+z)^{3(1+\alpha)(1+R)}}{l\widetilde{D}_{s}-(1-l)(1-\widetilde{D}_{s})(1+z)^{3(1+\alpha)(1+R)}}\right]^{\frac{1}{1+\alpha}}\right].    
\end{equation}\\

The parameters used in the above equation have the following forms:
$\widetilde{D}_{s}=\frac{1}{1+lP}$~, which satisfies the relation~$1-l<\widetilde{D}_{s}<1$.~Again, $\rho_{_{MCJG0}}$ represents the present energy density value of the MCJG having the expression $\rho_{_{MCJG0}}^{1+\alpha}=\frac{S}{(1+R)[l\widetilde{D}_{s}-(1-l)(1-\widetilde{D}_{s})]}$.\\

So, from the Eqn.(\ref{18}), we will get the equation of mass in terms of the redshift function $z$ as follows:

\begin{equation}\label{39}
M=\frac{M_0}{1+\frac{4\pi D_1 M_0}{3}\mathop{\mathlarger{\int}}_{\rho_{_{MCJG0}}}^{\rho_{_{MCJG}}}\frac{d\rho_{_{MCJG}}}{\sqrt{\frac{\kappa^2}{12}\left[\rho_{_{MCJG}}-6k\mu^4(1+z)^2-\frac{3k\kappa^2\mu^2(1+z)^4}{16}\right]}}}.
\end{equation}\\

where, the expression for $\rho_{_{MCJG}}$ is explicitly given by the Eqn.(\ref{density_MCJG}).\\

The above Eqn.(\ref{39}) is the desired mass equation of a 4-dimensional KS BH in the HL gravity scenario due to the accretion of MCJG. The corresponding constant $D_1$ has been obtained by Eqn.(\ref{ICE27}) in the Appendix.

\begin{figure}[H]
 \begin{subfigure}{.4\textwidth}
    \includegraphics[height=2.0in]{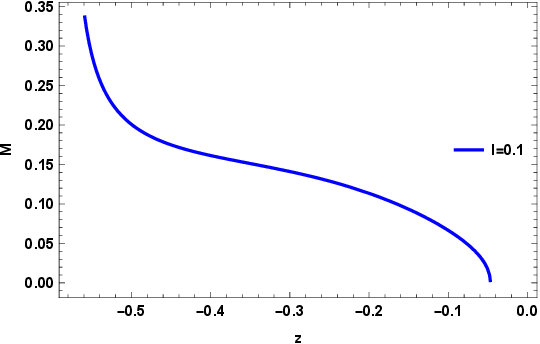}
    \subcaption{For the value of elliptic modulus: ($l)=0.1$~}
     \label{fig:f 6}
      \end{subfigure}
      \hfill
      \begin{subfigure}{.4\textwidth}
    \includegraphics[height=2.0in]{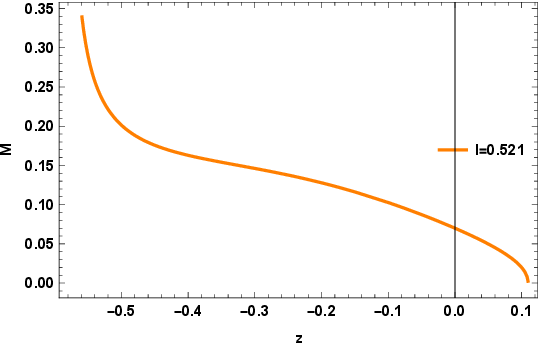}
    \subcaption{For the value of elliptic modulus: ($l)=0.521$~}
     \label{fig:f 6}
      \end{subfigure}
    \vfill
    \center
     \begin{subfigure}{.4\textwidth}
    \includegraphics[height=2.0in]{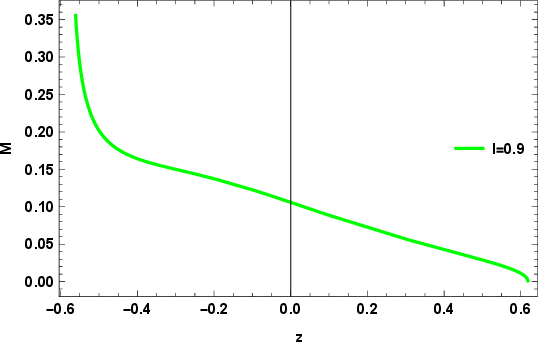}
    \subcaption{For the value of elliptic modulus: ($l)=0.9$~}
     \label{fig:f 6}
      \end{subfigure}
      \caption{The variations in BH mass $M$ against redshift $z$ due to the accretion of MCJG for different values of elliptic modulus $l$.[Other parameters: $\kappa=1$~,~$\mu=1$~,~$k=1$~,~$\omega_{_{ma}}=0.01$~,~$\Omega_{_{mo}}=0.21$~,~$\Omega_{_{MCJG0}}=0.71$~,~$H_0=72$~,~$M_0=10$~,~$\rho_\infty=0.1$~,~$\widetilde{D}_{s}=0.586$~,~$R=0.297$~,~$S=10$~and~$\alpha=0.066$]}
    \label{fig:my_label6}
\end{figure}

The depiction of the BH mass $M$ against the redshift $z$ has been shown through Fig.\ref{fig:my_label6} for different values of elliptic modulus($l) = 0.1~,~0.521~,~0.9$ respectively. It is quite evident from the above Fig.\ref{fig:my_label6} that due to the accretion of MCJG, there is an enhancement in the mass of a 4-dimensional KS BH, i.e., with the successive rise of the universe, black hole mass will also increase gradually because of MCJG accretion in HL gravity scenario.


\subsection{Modified Chaplygin-Abel Gas (MCAG)}\label{g}
As we all know, there are several types of elliptic function and hyperbolic functions are one of them. So, we can easily replace the hyperbolic function present in the generating function of the MCG model with the Abel elliptic function, which is a particular form of elliptic function, and then the consequent model is known as the modified Chaplygin-Abel gas (MCAG) model \cite{debnath2021roles}.\\
Therefore, for this model, the relation of pressure with energy density is given as follows \cite{debnath2021roles}:

\begin{equation}\label{40}
 p=\left[(b^2+2f^2)(1+R)-1\right]\rho-\frac{f^2S}{\rho^\alpha}-\frac{(b^2+f^2)(1+R)^2}{S}\rho^{2+\alpha} ,
\end{equation}

Here, both $R>0$ and $S>0$ are constants, $\alpha\in(0,1)$ is another constant, $b$ and $f$ are any two real numbers.Also, if we take $R=0$, then the above Eqn.(\ref{40}) will reduce to the equation of state for the generalized Chaplygin-Abel gas (GCAG) model \cite{debnath2021roles}.\\[0.5mm]
Now, the expression for the energy density will be given by \cite{debnath2021roles}:

\begin{equation}\label{41}
\rho=\left(\frac{S}{1+R}\right)^\frac{1}{1+\alpha}\left[\frac{a^{3b^2(1+\alpha)(1+R)}+f^2J}{a^{3b^2(1+\alpha)(1+R)}+(b^2+f^2)J}\right]^\frac{1}{1+\alpha}  ,
\end{equation}

With $J$ being a positive constant.\\
Again, the required energy density for MCAG accretion will be given by:

\begin{equation}\label{42}
  \rho_{_{MCAG}}=\rho_{_{m_0}}(1+z)^{3(1+\omega_{_{ma}})}+\left(\frac{S}{1+R}\right)^\frac{1}{1+\alpha}\left[\frac{(1+z)^{-3b^2(1+\alpha)(1+R)}+f^2J}{(1+z)^{-3b^2(1+\alpha)(1+R)}+(b^2+f^2)J}\right]^\frac{1}{1+\alpha} ,
\end{equation}\\

 Moreover, if we consider the Eqn.(\ref{density_dark matter}) and transform the above equation in terms of the dimensionless density parameter $\Omega_{_{MCAG0}}=\frac{\rho_{_{MCAG0}}}{3H^2_0}$~, then the above Eqn.(\ref{42}) will reduce to the following form \cite{debnath2021roles}:

\begin{multline}\label{density_MCAG}
 \rho_{_{MCAG}}=3H^2_0\left[\Omega_{_{m0}}(1+z)^{3(1+\omega_{_{ma}})}+\Omega_{_{MCAG0}}\left[f^2\widetilde{E}_{s}+(b^2+f^2)(1-\widetilde{E}_{s})\right]^{\frac{1}{1+\alpha}}\right.\\
\left.\left[\frac{\widetilde{E}_{s}+(1-\widetilde{E}_{s})(1+z)^{3b^2(1+\alpha)(1+R)}}{f^2\widetilde{E}_{s}+(b^2+f^2)(1-\widetilde{E}_{s})(1+z)^{3b^2(1+\alpha)(1+R)}}\right]^{\frac{1}{1+\alpha}}\right].   
\end{multline}\\

The parameters used in the above equation have the following forms:
$\widetilde{E}_{s}=\frac{1}{1+f^2J}$~,~which satisfies the relation~$0<\widetilde{E}_{s}<1$.~Also, the term $\rho_{_{MCAG0}}$ represents the present energy density value of MCAG having the expression $\rho_{_{MCAG0}}^{1+\alpha}=\frac{f^2S}{(1+R)[f^2\widetilde{E}_{s}+(b^2+f^2)(1-\widetilde{E}_{s})]}$.\\ 

Thus, from the Eqn.(\ref{18}), we will get the equation of mass in terms of the redshift function $z$ as follows:

\begin{equation}\label{43}
M=\frac{M_0}{1+\frac{4\pi D_1 M_0}{3}\mathop{\mathlarger{\int}}_{\rho_{_{MCAG0}}}^{\rho_{_{MCAG}}}\frac{d\rho_{_{MCAG}}}{\sqrt{\frac{\kappa^2}{12}\left[\rho_{_{MCAG}}-6k\mu^4(1+z)^2-\frac{3k\kappa^2\mu^2(1+z)^4}{16}\right]}}}.    
\end{equation}\\

where, the expression for $\rho_{_{MCAG}}$ is explicitly given by the Eqn.(\ref{density_MCAG}).\\ 

The above Eqn.(\ref{43}) is the desired mass equation of a 4-dimensional KS BH in the HL gravity scenario due to the accretion of MCAG. The corresponding constant $D_1$ has been obtained by Eqn.(\ref{ICE31}) in the Appendix.

\begin{figure}[H]
    \centering
    \includegraphics[width=0.45\linewidth]{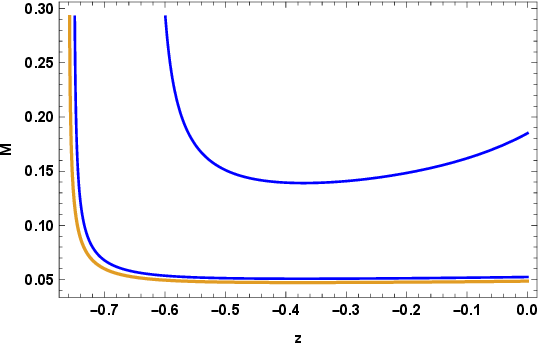}
    \caption{The variations in BH mass $M$ against redshift $z$ due to the accretion of MCAG for different values of $\bf{\alpha = 0.9,0.1,0.063}$ corresponding to down to up order of curves.[Other parameters: $\kappa=1$~,~$\mu=1$~,~$k=1$~,~$\omega_{_{ma}}=0.01$~,~$\Omega_{_{mo}}=0.21$~,~$\Omega_{_{MCAG0}}=0.72$~,~$H_0=72$~,~$M_0=10$~,~$\rho_\infty=0.1$~,~$R=0.276$~,~$S=100$~,~$b=0.172$~,~$f=0.392$~and~$\widetilde{E}_{s}=0.527$]}
\label{fig:my_label7}
\end{figure}

\begin{figure}[H]
 \begin{subfigure}{.4\textwidth}
    \includegraphics[height=2.0in]{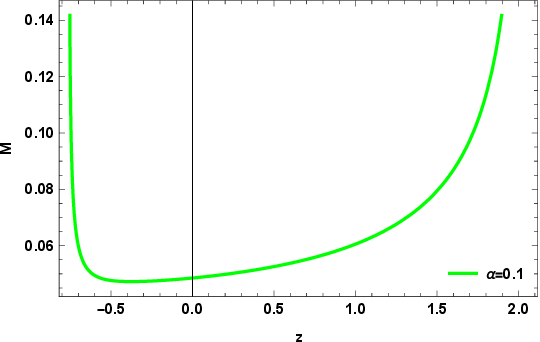}
    \subcaption{For the value of constant: ($\alpha=0.1$)}
     \label{fig:f 8}
      \end{subfigure}
      \hfill
      \begin{subfigure}{.4\textwidth}
    \includegraphics[height=2.0in]{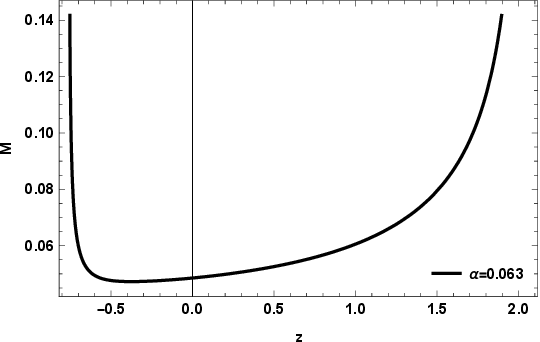}
    \subcaption{For the value of constant: ($\alpha=0.063$) }
     \label{fig:f 8}
      \end{subfigure}
    \vfill
    \center
     \begin{subfigure}{.4\textwidth}
    \includegraphics[height=2.0in]{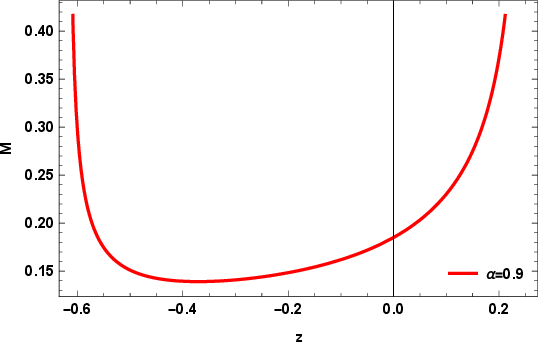}
    \subcaption{For the value of constant: ($\alpha=0.9$) }
     \label{fig:f 8}
      \end{subfigure}
      \caption{The variations in BH mass $M$ against redshift $z$ due to the accretion of MCAG for different values of constant $\alpha$.[Other parameters: $\kappa=1$~,~$\mu=1$~,~$k=1$~,~$\omega_{_{ma}}=0.01$~,~$\Omega_{_{mo}}=0.21$~,~$\Omega_{_{MCAG0}}=0.72$~,~$H_0=72$~,~$M_0=10$~,~$\rho_\infty=0.1$~,~$R=0.276$~,~$S=100$~,~$b=0.172$~,~$f=0.392$~and~$\widetilde{E}_{s}=0.527$ ]}
    \label{fig:my_label8}
\end{figure}

The depiction of the BH mass $M$ against the redshift $z$ has been shown through Fig.\ref{fig:my_label7} for different values of $\alpha = 0.9~,~0.1~,~0.063$ corresponding to the down to up order of the curves in a single framework. Also, in Fig.\ref{fig:my_label8}, the same has been plotted separately for different values of the constant $\alpha$ for better understanding of the change in mass with the redshift function. From both Fig.\ref{fig:my_label7} and Fig.\ref{fig:my_label8}, it is quite evident that due to the accretion of MCAG, there is an enhancement in the mass of a 4-dimensional KS BH, i.e., with the successive rise of the universe, the black hole mass will also increase gradually because of MCAG accretion in the HL gravity scenario.

\section{Discussions and Concluding Remarks:}\label{h}
From when the first idea of black holes was proposed until today, black holes have always been a very attractive topic for cosmologists. As mass is the most prominent property of a black hole, thus the accretion of mass onto black holes due to different kinds of dark energies has always been a very intriguing area of research among cosmologists. Today, observations on the changes in the mass of black holes resulting from various dark-energy accretions in different gravities have become quite a popular trend. Following this tendency, in this paper, we have analyzed the changes in the mass of a 4-dimensional Kehagias-Sfetsos(KS) black hole in the Ho{\v{r}}ava-Lifshitz(HL) gravity scenario due to the accretion of some special types of dark energy, mainly Chaplygin gas models.\\

We have first deduced the non-vanishing equations of motions in the case of HL gravity. Then, following the Babichev-Dokuchaev-Eroshenko method \cite{babichev2011perfect}, we have established the equation of mass in terms of the scale factor `$a$' for a 4-dimensional KS BH in the HL gravity scenario.\\

Then, we considered generalized cosmic Chaplygin gas(GCCG) to be a postulant of dark energy. The expression for the mass of a 4-dimensional KS BH due to the accretion of GCCG in HL gravity has been obtained and the graphical depiction of the BH mass $M$ against the redshift $z$ has been done in Fig.\ref{fig:my_label1} for different values of parameters $\widetilde{A}_{s}$, $\widetilde{A}$ and $w$. From Fig.\ref{fig:my_label1}, it is easily comprehensible that there is an enhancement in the mass of a 4-dimensional KS BH, i.e., with the successive rise of the universe, the black hole mass will also increase gradually because of GCCG accretion in the HL gravity scenario.\\

After that, we used variable modified Chaplygin gas (VMCG) as a postulant for dark energy. The expression for the mass of a 4-dimensional KS BH due to the accretion of VMCG in HL gravity has been obtained, and the graphical depiction of the BH mass $M$ against the redshift $z$ has been done in Fig.\ref{fig:my_label2} for different values of $\alpha$. Also, in Fig.\ref{fig:my_label3}, we have shown the same concerning different values of the parameter $\widetilde{B}_{s}$ and constants $P$~,~$m$. From Fig.\ref{fig:my_label2} and Fig.\ref{fig:my_label3}, it is easily
comprehensible that there is an enhancement in the mass of a 4-dimensional KS BH, i.e., with the successive rise of the universe, the black hole mass will also increase gradually because of VMCG accretion in the HL gravity scenario.\\

Then, we considered new variable modified Chaplygin gas(NVMCG) as a postulant of dark energy. The expression for the mass of a 4-dimensional KS BH due to the accretion of NVMCG in HL gravity has been obtained, and the graphical depiction of the black hole mass $M$ against the redshift $z$ has been done in Fig.\ref{fig:my_label4} for different values of $\alpha$. Again, in Fig.\ref{fig:my_label5}, we have shown the same concerning different values of the parameter~$\widetilde{C}_{s}$~and constants~$P$~and~$\alpha$. From Fig.\ref{fig:my_label4} and Fig.\ref{fig:my_label5}, it is easily comprehensible that there is an enhancement in the mass of a 4-dimensional KS BH, i.e., with the successive rise of the universe, the black hole mass will also increase gradually because of NVMCG accretion in the HL gravity scenario.\\

Again, we have taken modified Chaplygin-Jacobi gas(MCJG) as a postulant of dark energy. The expression for the mass of a 4-dimensional KS BH due to the accretion of MCJG in HL gravity has been obtained, and the graphical depiction of the black hole mass $M$ against the redshift $z$ has been done in Fig.\ref{fig:my_label6} for different values of the elliptic modulus ($l$). From Fig.\ref{fig:my_label6}, it is easily comprehensible that there is an enhancement in the mass of a 4-dimensional KS BH, i.e., with the successive rise of the universe, the black hole mass will also increase gradually because of MCJG accretion in the HL gravity scenario.\\

Lastly, we have considered modified Chaplygin-Abel gas(MCAG) as a postulant of dark energy. The expression for the mass of a 4-dimensional KS BH due to the accretion of MCAG in HL gravity has been obtained, and the graphical depiction of the black hole mass $M$ against the redshift $z$ has been done in Fig.\ref{fig:my_label7} for different values of $\alpha$ in a single frame, as well as in Fig.\ref{fig:my_label8} we have shown the same for different values of $\alpha$ separately for better visualization. From both Fig.\ref{fig:my_label7} and Fig.\ref{fig:my_label8}, it is easily comprehensible that there is an enhancement in the mass of a 4-dimensional KS BH, i.e., with the successive rise of the universe, the black hole mass will also increase gradually because of the MCAG accretion in the HL gravity scenario.\\

So, in general, we can conclude that the mass of a 4-dimensional KS BH will increase gradually due to the accretion of dark energy candidates like GCCG, VMCG, NVMCG, MCJG, and MCAG in the case of HL gravity-induced universe.


\section*{\normalsize\bf{ACKNOWLEDGEMENTS}}
PM is thankful to IIEST, Shibpur, India, for providing an Institute Fellowship (SRF). The facility and assistance under visiting associateship were provided by the IUCAA, Pune, India, for which AP is grateful.\\\\


{\Large\bf Appendix:}

\begin{itemize}
    \item {\textbf{Calculation of constant $D_1$:}}
To calculate the dimensionless, positive constant $D_1$ let us consider the fluid flow $u=\frac{dr}{ds}<0$ and the dimensionless radius $x\equiv\frac{r}{M}$ for accretion, then Eqn.(\ref{8}) provide us with the expression of dimensionless, positive constant $D_1$ as follows:

\begin{equation}\label{ICE1}
 ux^2exp\left[\int_{\rho_\infty}^{\rho}{\frac{d\rho\prime}{\rho\prime+p(\rho\prime)}}\right]=-D_1,   
\end{equation}

Now, let us consider a formal auxiliary function $n$, defined by

\begin{equation}\label{ICE2}
 \frac{d\rho}{\rho+p}=\frac{dn}{n},   
\end{equation}

valid for any given equation of state $p=p(\rho)$. Here, $n$ represents the particle number density when the medium has only individual conserved particles \cite{babichev2005accretion}. A solution for the above-mentioned concentration $n$ is given by:

\begin{equation}\label{ICE3}
\frac{n}{n_\infty}\equiv exp\left[\int_{\rho_\infty}^{\rho}{\frac{d\rho\prime}{\rho\prime+p(\rho\prime)}}\right].    
\end{equation}

The term $n_\infty$ is called the concentration of dark energy, introduced for our own convenience \cite{babichev2005accretion}.

Again, from Eqn.(\ref{7}) we have the following relation:

\begin{equation}\label{ICE4}
ux^2(\rho+p)\left(\sqrt{u^2+\mathcal{A}}\right)=D_0,    
\end{equation}

and Eqn.(\ref{9}) converts into: 

\begin{equation}\label{ICE5}
\frac{(\rho+p)}{n}\left(\sqrt{u^2+\mathcal{A}}\right)=D_2,    
\end{equation}

where, $D_2 $ is given by the relation: $D_2=-\frac{D_0}{D_1}=\frac{\rho_\infty+p(\rho_\infty)}{n_\infty}$.\\[1.5mm]
The value of $D_1$, which is nothing but an energy flux constant, can be obtained by evaluating the fluid parameters at any critical point. It will also provide us with the continuity of the solution between a horizon and infinity, i.e., the fluid flow should cross the critical point smoothly \cite{michel1972accretion}. In this regard, let us consider a parameter having the same dimension as the velocity given by:

\begin{equation}\label{ICE6}
 V^2=\frac{n}{\rho+p}\frac{d(\rho+p)}{dn}-1,  
\end{equation}

By Eqn.(\ref{ICE2}), we can easily conclude that $V^2$ is equivalent to the effective sound speed in the considered medium, denoted by the symbol $c^2_s(\rho)$. So, it can be written in the following form:

\begin{equation}\label{ICE7}
 V^2=c^2_s(\rho)=\frac{\partial p}{\partial \rho},   
\end{equation}

Now, differentiating Eqn.(\ref{ICE1}) and Eqn.(\ref{ICE5}) at the same time using Eqn.(\ref{ICE3}) we deduced the following relation:

\begin{equation}\label{ICE8}
\frac{du}{u}\left[V^2-\frac{u^2}{u^2+\mathcal{A}}\right]+\frac{dx}{x}\left[2V^2-\frac{x\mathcal{A}^\prime}{2(u^2+\mathcal{A})}\right]=0.  
\end{equation}

Taking both square brackets equal to zero (to obtain the single-valued solutions) in the above Eqn.(\ref{ICE8}), we have the required relationship between the parameters at the critical point as follows:

\begin{equation}\label{ICE9}
u^2_c=\frac{x_c\mathcal{A}^\prime_c}{4}~~~~~~and~~~~~~V^2_c=\frac{x_c\mathcal{A}^\prime_c}{x_c\mathcal{A}^\prime_c+4\mathcal{A}_c},    
\end{equation}

By subscript ``$c$", we have expressed the critical values and $u_c$ represents the critical flow speed at any critical point $x=x_c$. Also, the expressions for $\mathcal{A}_c$ and $\mathcal{A}^\prime_c$ are given in respective manner as follows:

\begin{align}\label{metric_function}
\mathcal{A}_c=1+\omega r^2_c\left(1-\sqrt{1+\frac{4}{\omega x_cr^2_c}}\right),\\
\mathcal{A}^\prime_c=2\omega\frac{r^2_c}{x_c}\left(1-\sqrt{1+\frac{4}{\omega x_cr^2_c}}\right)+\frac{6}{x^2_c\sqrt{1+\frac{4}{\omega x_cr^2_c}}}.
\end{align}

Again, $V^2_c=c^2_s(\rho_c)$, the term $\rho_c$ represents the energy density at any critical point $x=x_c$. \\

Now, from Eqn.(\ref{9}), using  Eqn.(\ref{ICE9}) we established the following relation for the critical point:

\begin{equation}\label{ICE10}
\frac{\rho_c+p(\rho_c)}{\rho_\infty+p(\rho_\infty)}=2\left[x_c\mathcal{A}^\prime_c+4\mathcal{A}_c\right]^{-1/2}exp\left[\int_{\rho_\infty}^{\rho_c}{\frac{d\rho\prime}{\rho\prime+p(\rho\prime)}}\right],    
\end{equation}

Using the above equation, we can easily obtain the density at any critical point given as $\rho_c=\rho(x_c)$. Again, for any particular value of $\rho_c$, Eqn.(\ref{ICE9}) will simultaneously give us our desired values of $x_c$ and $u_c$.\\

Finally from Eqn.(\ref{ICE1}) we have our desired expression of the constant $D_1$ in the following form:

\begin{equation}\label{ICE11}
 D_1=\frac{\left(x^3_c\mathcal{A}^\prime_c\right)^{1/2}}{2}exp\left[\int_{\rho_\infty}^{\rho_c}{\frac{d\rho\prime}{\rho\prime+p(\rho\prime)}}\right].   
\end{equation}

Hence, for different dark energy models, particularly in our case, for different Chaplygin gas models, we have to calculate the R.H.S. of Eqn.(\ref{ICE11}) to get the corresponding constant $D_1$ in the background of Ho{\v{r}}ava-Lifshitz (HL) gravity. Also, we can easily recover the form of the corresponding $D_1$ for a Schwarzschild black hole in the case of Chaplygin gas accretion, as discussed in Ref. \cite{babichev2005accretion}, from Eqn.(\ref{ICE11}). This confirms the validity of the above Eqn.(\ref{ICE11}).

\end{itemize}

\begin{itemize}
 \item{\textbf{Generalized Cosmic Chaplygin Gas (GCCG):}} 
To obtain the constant $D_1$ for GCCG model, at first using Eqn.(\ref{19}), we have the following expression:
\begin{equation}\label{ICE12}
exp\left[\int_{\rho_\infty}^{\rho}{\frac{d\rho\prime}{\rho\prime+p(\rho\prime)}}\right]=\left[\frac{\left(\rho^{\alpha+1}-G\right)^{w+1}-1}{\left(\rho^{\alpha+1}_\infty-G\right)^{w+1}-1}\right]^{\frac{1}{(\alpha+1)(w+1)}},   
\end{equation}

Taking Eqn.(\ref{19}) and Eqn.(\ref{ICE9}) simultaneously, we get the critical points as follows:

\begin{multline}\label{ICE13}
u^2_c=\mathcal{A}_c\left[\frac{\alpha\rho^{-\alpha-1}_c\{G+(\rho^{\alpha+1}_c-G)^{-w}\}+w(\alpha+1)(\rho^{\alpha+1}_c-G)^{-w-1}}{1-\alpha\rho^{-\alpha-1}_c\{G+(\rho^{\alpha+1}_c-G)^{-w}\}-w(\alpha+1)(\rho^{\alpha+1}_c-G)^{-w-1}}\right],\\[1.5mm] 
x_c=\frac{4\mathcal{A}_c}{\mathcal{A}^\prime_c}\left[\frac{\alpha\rho^{-\alpha-1}_c\{G+(\rho^{\alpha+1}_c-G)^{-w}\}+w(\alpha+1)(\rho^{\alpha+1}_c-G)^{-w-1}}{1-\alpha\rho^{-\alpha-1}_c\{G+(\rho^{\alpha+1}_c-G)^{-w}\}-w(\alpha+1)(\rho^{\alpha+1}_c-G)^{-w-1}}\right].       
\end{multline}

Also, by Eqn.(\ref{ICE10}) we get the following relation for $\rho_c$ as:

\begin{equation}\label{ICE14}
\left[\frac{\rho_c-\rho^{-\alpha}_c\{G+(\rho^{\alpha+1}_c-G)^{-w}\}}{\rho_\infty-\rho^{-\alpha}_\infty\{G+(\rho^{\alpha+1}_\infty-G)^{-w}\}}\right]=2\left[x_c\mathcal{A}^\prime_c+4\mathcal{A}_c\right]^{-1/2}\left[\frac{\left(\rho^{\alpha+1}_c-G\right)^{w+1}-1}{\left(\rho^{\alpha+1}_\infty-G\right)^{w+1}-1}\right]^{\frac{1}{(\alpha+1)(w+1)}},
\end{equation}

Hence, our desired form of the constant $D_1$ due to GCCG accretion is given by Eqn.(\ref{ICE11}) as follows:

\begin{equation}\label{ICE15}
D_1=\frac{4}{\mathcal{A}^\prime_c}\left[\mathcal{A}_c~\left\{\frac{\alpha\rho^{-\alpha-1}_c\{G+(\rho^{\alpha+1}_c-G)^{-w}\}+w(\alpha+1)(\rho^{\alpha+1}_c-G)^{-w-1}}{1-\alpha\rho^{-\alpha-1}_c\{G+(\rho^{\alpha+1}_c-G)^{-w}\}-w(\alpha+1)(\rho^{\alpha+1}_c-G)^{-w-1}}\right\}\right]^{3/2}\left[\frac{\left(\rho^{\alpha+1}_c-G\right)^{w+1}-1}{\left(\rho^{\alpha+1}_\infty-G\right)^{w+1}-1}\right]^\frac{1}{(\alpha+1)(w+1)}.      
\end{equation}

\end{itemize}

\begin{itemize}
    \item \textbf{Variable Modified Chaplygin Gas (VMCG):}

For VMCG model, using both the Eqn.(\ref{25}) and Eqn.(\ref{26}) together, we have the following expression:
\begin{equation}\label{ICE16}
exp\left[\int_{\rho_\infty}^{\rho}{\frac{d\rho\prime}{\rho\prime+p(\rho\prime)}}\right]=\left[\frac{\rho^{\alpha+1}(P+1)-Qa^{-m}}{\rho^{\alpha+1}_\infty(P+1)-Qa^{-m}_\infty}\right]^\frac{1}{(\alpha+1)(P+1)},    
\end{equation}

Here, $a_\infty$ is the cosmological scale factor at infinity, and its relationship with $\rho_\infty$ is given by Eqn.(\ref{27}).\\[0.5mm]
Again, considering Eqn.(\ref{25}) and Eqn.(\ref{26}) in Eqn.(\ref{ICE9}) we get the critical points as:

\begin{equation}\label{ICE17}
u^2_c=\mathcal{A}_c\left[\frac{P+\alpha Qa^{-m}_c\rho^{-\alpha-1}_c}{1-P-\alpha Qa^{-m}_c\rho^{-\alpha-1}_c}\right],~~~~and~~~~
x_c=\frac{4\mathcal{A}_c}{\mathcal{A}^\prime_c}\left[\frac{P+\alpha Qa^{-m}_c\rho^{-\alpha-1}_c}{1-P-\alpha Qa^{-m}_c\rho^{-\alpha-1}_c}\right].
\end{equation}

where $a_c$ is the cosmological scale factor at the critical point, and the relationship between $\rho_c$ and $a_c$ is given by the Eqn.(\ref{27}).\\
To get the relation for $\rho_c$, using the above Eqn.(\ref{ICE10}) we have:
\begin{equation}\label{ICE18}
 \left[\frac{\rho_c(P+1)-Qa^{-m}_c\rho^{-\alpha}_c}{\rho_\infty(P+1)-Qa^{-m}_\infty\rho^{-\alpha}_\infty}\right]=2\left[x_c\mathcal{A}^\prime_c+4\mathcal{A}_c\right]^{-1/2}\left[\frac{\rho^{\alpha+1}_c(P+1)-Qa^{-m}_c}{\rho^{\alpha+1}_\infty(P+1)-Qa^{-m}_\infty}\right]^\frac{1}{(\alpha+1)(P+1)},
 \end{equation}

 Finally, using the Eqn.(\ref{ICE11}) we have the required expression of the constant $D_1$ for the VMCG accretion model in the following form:

 \begin{equation}\label{ICE19}
 D_1=\frac{4}{\mathcal{A}^\prime_c}\left[\mathcal{A}_c~\left\{\frac{P+\alpha Qa^{-m}_c\rho^{-\alpha-1}_c}{1-P-\alpha Qa^{-m}_c\rho^{-\alpha-1}_c}\right\}\right]^{3/2}\left[\frac{\rho^{\alpha+1}_c(P+1)-Qa^{-m}_c}{\rho^{\alpha+1}_\infty(P+1)-Qa^{-m}_\infty}\right]^\frac{1}{(\alpha+1)(P+1)}.    
 \end{equation}

\end{itemize}

\begin{itemize}
    \item{\textbf{New Variable Modified Chaplygin Gas (NVMCG):}}

To obtain the constant $D_1$ for NVMCG model, using Eqn.(\ref{30}), Eqn.(\ref{31}) and Eqn.(\ref{32}) together we have the following expression:
\begin{equation}\label{ICE20}
exp\left[\int_{\rho_\infty}^{\rho}{\frac{d\rho\prime}{\rho\prime+p(\rho\prime)}}\right]=\left[\frac{\rho^{\alpha+1}\left(Pa^{-j}+1\right)-Qa^{-i}}{\rho^{\alpha+1}_\infty\left(Pa^{-j}_\infty+1\right)-Qa^{-i}_\infty}\right]^\frac{1}{(\alpha+1)(Pa^{-j}+1)},    
\end{equation}

Here, $a_\infty$ is the cosmological scale factor at infinity, and its relationship with $\rho_\infty$ is given by Eqn.(\ref{33}).\\

Again, considering Eqn.(\ref{25}) and Eqn.(\ref{26}) in Eqn.(\ref{ICE9}) we get the critical points as:

\begin{equation}\label{ICE21}
  u^2_c=\mathcal{A}_c\left[\frac{Pa^{-j}_c+\alpha Qa^{-i}_c\rho^{-\alpha-1}_c}{1-Pa^{-j}_c-\alpha Qa^{-i}_c\rho^{-\alpha-1}_c}\right],~~~~and~~~~x_c=\frac{4\mathcal{A}_c}{\mathcal{A}^\prime_c}\left[\frac{Pa^{-j}_c+\alpha Qa^{-i}_c\rho^{-\alpha-1}_c}{1-Pa^{-j}_c-\alpha Qa^{-i}_c\rho^{-\alpha-1}_c}\right].   
\end{equation}

The relation between cosmological scale factor $a_c$ and critical density $\rho_c$ is given by Eqn.(\ref{33}).\\[0.5mm]
Again, considering Eqn.(\ref{ICE10}) we have the following relation for $\rho_c$ as:

\begin{equation}\label{ICE22}
\left[\frac{\rho_c(Pa^{-j}_c+1)-Qa^{-i}_c\rho^{-\alpha}_c}{\rho_\infty(Pa^{-j}_\infty+1)-Qa^{-i}_\infty\rho^{-\alpha}_\infty}\right]=2\left[x_c\mathcal{A}^\prime_c+4\mathcal{A}_c\right]^{-1/2}\left[\frac{\rho^{\alpha+1}_c\left(Pa^{-j}_c+1\right)-Qa^{-i}_c}{\rho^{\alpha+1}_\infty\left(Pa^{-j}_\infty+1\right)-Qa^{-i}_\infty}\right]^\frac{1}{(\alpha+1)(Pa^{-j}_c+1)},            
\end{equation}

Thus, our desired form of the constant $D_1$ due to the accretion of NVMCG is given by Eqn.(\ref{ICE11}) as follows:

\begin{equation}\label{ICE23}
D_1=\frac{4}{\mathcal{A}^\prime_c}\left[\mathcal{A}_c~\left\{\frac{Pa^{-j}_c+\alpha Qa^{-i}_c\rho^{-\alpha-1}_c}{1-Pa^{-j}_c-\alpha Qa^{-i}_c\rho^{-\alpha-1}_c}\right\}\right]^{3/2}\left[\frac{\rho^{\alpha+1}_c\left(Pa^{-j}_c+1\right)-Qa^{-i}_c}{\rho^{\alpha+1}_\infty\left(Pa^{-j}_\infty+1\right)-Qa^{-i}_\infty}\right]^\frac{1}{(\alpha+1)(Pa^{-j}_c+1)}.    
\end{equation}
\end{itemize}

\begin{itemize}
    \item {\textbf{Modified Chaplygin-Jacobi Gas (MCJG):}}
To obtain the constant $D_1$ for the MCJG model, first using Eqn.(\ref{36}), we have the following expression:

\begin{equation}\label{ICE24}
exp\left[\int_{\rho_\infty}^{\rho}{\frac{d\rho\prime}{\rho\prime+p(\rho\prime)}}\right]=\left[\frac{\{(1+R)\rho^{\alpha+1}-S\}\{(1-l)(1+R)\rho^{\alpha+1}_\infty+S\}}{\{(1+R)\rho^{\alpha+1}_\infty-S\}\{(1-l)(1+R)\rho^{\alpha+1}+S\}}\right]^\frac{1}{(\alpha+1)(R+1)},    
\end{equation}

The required critical points are given by using Eqn.(\ref{36}) in Eqn.(\ref{ICE9}) as follows:

\begin{multline}\label{ICE25}
u^2_c=\mathcal{A}_c\left[\frac{\{(\alpha+2)(1-l)(1+R)^2\}\rho^{\alpha+1}_c+\alpha lS^2\rho^{-\alpha-1}_c+\{(2l-1)(1+R)-1\}S}{2S-\left\{(\alpha+2)(1-l)(1+R)^2\right\}\rho^{\alpha+1}_c-\alpha lS^2\rho^{-\alpha-1}_c-(2l-1)(1+R)S}\right],\\
x_c=\frac{4\mathcal{A}_c}{\mathcal{A}^\prime_c}\left[\frac{\{(\alpha+2)(1-l)(1+R)^2\}\rho^{\alpha+1}_c+\alpha lS^2\rho^{-\alpha-1}_c+\{(2l-1)(1+R)-1\}S}{2S-\left\{(\alpha+2)(1-l)(1+R)^2\right\}\rho^{\alpha+1}_c-\alpha lS^2\rho^{-\alpha-1}_c-(2l-1)(1+R)S}\right].
\end{multline}

Also,by the relation given in Eqn.(\ref{ICE10}) we have:

\begin{multline}\label{ICE26}
\left[\frac{(1-l)(1+R)^2\rho^{\alpha+2}_c+(2l-1)(1+R)S\rho_c-lS^2\rho^{-\alpha}_c}{(1-l)(1+R)^2\rho^{\alpha+2}_\infty+(2l-1)(1+R)S\rho_\infty-lS^2\rho^{-\alpha}_\infty}\right]\\=2\left[x_c\mathcal{A}^\prime_c+4\mathcal{A}_c\right]^{-1/2}
\left[\frac{\{(1+R)\rho^{\alpha+1}_c-S\}\{(1-l)(1+R)\rho^{\alpha+1}_\infty+S\}}{\{(1+R)\rho^{\alpha+1}_\infty-S\}\{(1-l)(1+R)\rho^{\alpha+1}_c+S\}}\right]^\frac{1}{(\alpha+1)(R+1)},    
\end{multline}

Finally, the required expression of the constant $D_1$ for the MCJG accretion model is given by Eqn.(\ref{ICE11}) as follows:

\begin{multline}\label{ICE27}
D_1=\frac{4}{\mathcal{A}^\prime_c}\left[\mathcal{A}_c~\left\{\frac{\{(\alpha+2)(1-l)(1+R)^2\}\rho^{\alpha+1}_c+\alpha lS^2\rho^{-\alpha-1}_c+\{(2l-1)(1+R)-1\}S}{2S-\left\{(\alpha+2)(1-l)(1+R)^2\right\}\rho^{\alpha+1}_c-\alpha lS^2\rho^{-\alpha-1}_c-(2l-1)(1+R)S}\right\}\right]^{3/2}\\
\left[\frac{\{(1+R)\rho^{\alpha+1}_c-S\}\{(1-l)(1+R)\rho^{\alpha+1}_\infty+S\}}{\{(1+R)\rho^{\alpha+1}_\infty-S\}\{(1-l)(1+R)\rho^{\alpha+1}_c+S\}}\right] ^\frac{1}{(\alpha+1)(R+1)}.  
\end{multline}

\end{itemize}

\begin{itemize}
    \item {\textbf{Modified Chaplygin-Abel Gas (MCAG):}}

To obtain the constant $D_1$ for MCAG model, using Eqn.(\ref{40}) we get the following expression:
\begin{equation}\label{ICE28}
exp\left[\int_{\rho_\infty}^{\rho}{\frac{d\rho\prime}{\rho\prime+p(\rho\prime)}}\right]=\left[\frac{\{(b^2+f^2)(1+R)\rho^{\alpha+1}-f^2S\}\{(b^2+f^2)(1+R)\rho^{\alpha+1}_\infty-(b^2+f^2)S\}}{\{(b^2+f^2)(1+R)\rho^{\alpha+1}-(b^2+f^2)S\}\{(b^2+f^2)(1+R)\rho^{\alpha+1}_\infty-f^2S\}}\right]^\frac{1}{b^2(\alpha+1)(R+1)},    
\end{equation}

Again, for critical points, using  Eqn.(\ref{40}) and Eqn.(\ref{ICE9}) together we have;

\begin{multline}\label{ICE29}
u^2_c=\mathcal{A}_c\left[\frac{\{(b^2+2f^2)(1+R)-1\}S+\alpha f^2S^2\rho^{-\alpha-1}_c-\{(\alpha+2)(b^2+f^2)(1+R)^2\}\rho^{\alpha+1}_c}{2S-(b^2+2f^2)(1+R)S-\alpha f^2S^2\rho^{-\alpha-1}_c+\{(\alpha+2)(b^2+f^2)(1+R)^2\}\rho^{\alpha+1}_c}\right],\\[1.5mm]
x_c=\frac{4\mathcal{A}_c}{\mathcal{A}^\prime_c}\left[\frac{\{(b^2+2f^2)(1+R)-1\}S+\alpha f^2S^2\rho^{-\alpha-1}_c-\{(\alpha+2)(b^2+f^2)(1+R)^2\}\rho^{\alpha+1}_c}{2S-(b^2+2f^2)(1+R)S-\alpha f^2S^2\rho^{-\alpha-1}_c+\{(\alpha+2)(b^2+f^2)(1+R)^2\}\rho^{\alpha+1}_c}\right].
\end{multline}

Again, utilizing Eqn.(\ref{ICE10}) we have the following relation for $\rho_c$ as:
\begin{multline}\label{ICE30}
\left[\frac{(b^2+f^2)(1+R)^2\rho^{\alpha+2}_c-S(b^2+2f^2)(1+R)\rho_c+f^2S^2\rho^{-\alpha}_c}{(b^2+f^2)(1+R)^2\rho^{\alpha+2}_\infty-S(b^2+2f^2)(1+R)\rho_\infty+f^2S^2\rho^{-\alpha}_\infty}\right]=2\left[x_c\mathcal{A}^\prime_c+4\mathcal{A}_c\right]^{-1/2}\\
\left[\frac{\{(b^2+f^2)(1+R)\rho^{\alpha+1}_c-f^2S\}\{(b^2+f^2)(1+R)\rho^{\alpha+1}_\infty-(b^2+f^2)S\}}{\{(b^2+f^2)(1+R)\rho^{\alpha+1}_c-(b^2+f^2)S\}\{(b^2+f^2)(1+R)\rho^{\alpha+1}_\infty-f^2S\}}\right]^\frac{1}{b^2(\alpha+1)(R+1)},
\end{multline}

Hence, the desired form of constant $D_1$ due to the MCAG accretion is given by utilizing the Eqn.(\ref{ICE11}) in the following form:

\begin{multline}\label{ICE31}
D_1=\frac{4}{\mathcal{A}^\prime_c}\left[\mathcal{A}_c~\left\{\frac{\{(b^2+2f^2)(1+R)-1\}S+\alpha f^2S^2\rho^{-\alpha-1}_c-\{(\alpha+2)(b^2+f^2)(1+R)^2\}\rho^{\alpha+1}_c}{2S-(b^2+2f^2)(1+R)S-\alpha f^2S^2\rho^{-\alpha-1}_c+\{(\alpha+2)(b^2+f^2)(1+R)^2\}\rho^{\alpha+1}_c}\right\}\right]^{3/2}\\ 
\left[\frac{\{(b^2+f^2)(1+R)\rho^{\alpha+1}_c-f^2S\}\{(b^2+f^2)(1+R)\rho^{\alpha+1}_\infty-(b^2+f^2)S\}}{\{(b^2+f^2)(1+R)\rho^{\alpha+1}_c-(b^2+f^2)S\}\{(b^2+f^2)(1+R)\rho^{\alpha+1}_\infty-f^2S\}}\right]^\frac{1}{b^2(\alpha+1)(R+1)}.
\end{multline}

\end{itemize}


\printbibliography

@article{bhattacharya2011thermodynamical,
  title={\href{https://doi.org/10.1142/S0218271811019323}{Thermodynamical Laws in Ho{\v{r}}ava--Lifshitz Gravity}},
  author={Bhattacharya, Samarpita and Debnath, Ujjal},
  journal={International Journal of Modern Physics D},
  volume={20},
  number={07},
  pages={1191--1204},
  year={2011},
  publisher={World Scientific}
}

@article{abbas2014phantom,
  title={\href{
https://doi.org/10.1007/s11433-013-5306-z
}{Phantom energy accretion onto a black hole in Ho{\v{r}}ava-Lifshitz gravity}},
  author={Abbas, G},
  journal={Science China Physics, Mechanics and Astronomy},
  volume={57},
  pages={604--607},
  year={2014},
  publisher={Springer}
}

@article{liu2011logarithmic,
  title={\href{https://doi.org/10.1016/j.physletb.2011.04.006}{Logarithmic entropy of Kehagias--Sfetsos black hole with self-gravitation in asymptotically flat IR modified Ho{\v{r}}ava gravity}},
  author={Liu, Molin and Lu, Junwang},
  journal={Physics Letters B},
  volume={699},
  number={4},
  pages={296--300},
  year={2011},
  publisher={Elsevier}
}

@article{john2013accretion,
  title={\href{https://doi.org/10.1103/PhysRevD.88.104005}{Accretion onto a higher dimensional black hole}},
  author={John, Anslyn J and Ghosh, Sushant G and Maharaj, Sunil D},
  journal={Physical Review D},
  volume={88},
  number={10},
  pages={104005},
  year={2013},
  publisher={APS}
}

@article{debnath2015accretions,
  title={\href{https://doi.org/10.1007/s10509-015-2552-8 }{Accretions of dark matter and dark energy onto (n+ 2)-dimensional Schwarzschild black hole and Morris-Thorne wormhole}},
  author={Debnath, Ujjal},
  journal={Astrophysics and Space Science},
  volume={360},
  number={2},
  pages={40},
  year={2015},
  publisher={Springer}
}

@article{babichev2011perfect,
  title={\href{https://dx.doi.org/10.1134/S1063776111040157}{Perfect fluid and scalar field in the Reissner-Nordstr{\"o}m metric}},
  author={Babichev, EO and Dokuchaev, VI and Eroshenko, Yu N},
  journal={Journal of Experimental and Theoretical Physics},
  volume={112},
  pages={784--793},
  year={2011},
  publisher={Springer}
}

@article{gonzalez2003you,
  title={\href{https://doi.org/10.1103/PhysRevD.68.021303}{You need not be afraid of phantom energy}},
  author={Gonz{\'a}lez-D{\'\i}az, Pedro F},
  journal={Physical Review D},
  volume={68},
  number={2},
  pages={021303},
  year={2003},
  publisher={APS}
}

@article{chakraborty2007generalized,
  title={\href{
https://doi.org/10.48550/arXiv.0711.0079
}{Generalized cosmic Chaplygin gas model with or without interaction}},
  author={Chakraborty, Writambhara and Debnath, Ujjal and Chakraborty, Subenoy},
  journal={arXiv preprint arXiv:0711.0079},
  year={2007}
}

@article{debnath2007variable,
  title={\href{ 
https://doi.org/10.1007/s10509-007-9690-6 }{Variable modified Chaplygin gas and accelerating universe}},
  author={Debnath, Ujjal},
  journal={Astrophysics and Space Science},
  volume={312},
  number={3-4},
  pages={295--299},
  year={2007},
  publisher={Springer}
}

@article{chakraborty2010new,
  title={\href{https://doi.org/10.1134/S0202289310030059 }{A new variable modified Chaplygin gas model interacting with a scalar field}},
  author={Chakraborty, Writambhara and Debnath, Ujjal},
  journal={Gravitation and Cosmology},
  volume={16},
  number={3},
  pages={223--227},
  year={2010},
  publisher={Springer}
}

@article{debnath2004role,
  title={\href{https://doi.org/10.1088/0264-9381/21/23/019
}{Role of modified Chaplygin gas in accelerated universe}},
  author={Debnath, Ujjal and Banerjee, Asit and Chakraborty, Subenoy},
  journal={Classical and Quantum Gravity},
  volume={21},
  number={23},
  pages={5609},
  year={2004},
  publisher={IOP Publishing}
}

@article{villanueva2015generalized,
  title={\href{https://doi.org/10.1088/1475-7516/2015/07/045}{The generalized Chaplygin-Jacobi gas}},
  author={Villanueva, JR},
  journal={Journal of Cosmology and Astroparticle Physics},
  volume={2015},
  number={07},
  pages={045},
  year={2015},
  publisher={IOP Publishing}
}

@article{debnath2021roles,
  title={\href{https://doi.org/10.1142/S0217751X21502456}{Roles of modified Chaplygin--Jacobi and Chaplygin--Abel gases in FRW universe}},
  author={Debnath, Ujjal},
  journal={International Journal of Modern Physics A},
  volume={36},
  number={33},
  pages={2150245},
  year={2021},
  publisher={World Scientific}
}

@article{kamenshchik2001alternative,
  title={\href{https://ui.adsabs.harvard.edu/link_gateway/2001PhLB..511..265K/doi:10.1016/S0370-2693(01)00571-8}{An alternative to quintessence}},
  author={Kamenshchik, Alexander and Moschella, Ugo and Pasquier, Vincent},
  journal={Physics Letters B},
  volume={511},
  number={2-4},
  pages={265--268},
  year={2001},
  publisher={Elsevier}
}

@article{rudra2013role,
  title={\href{https://doi.org/10.1142/S0217732313501022}{Role of generalized cosmic Chaplygin gas in accelerating universe: a field theoretical prescription}},
  author={Rudra, Prabir},
  journal={Modern Physics Letters A},
  volume={28},
  number={22},
  pages={1350102},
  year={2013},
  publisher={World Scientific}
}

@article{gorini2003can,
  title={\href{https://doi.org/10.1103/PhysRevD.67.063509}{Can the Chaplygin gas be a plausible model for dark energy?}},
  author={Gorini, Vittorio and Kamenshchik, Alexander and Moschella, Ugo},
  journal={Physical Review D},
  volume={67},
  number={6},
  pages={063509},
  year={2003},
  publisher={APS}
}

@article{benaoum2012modified,
  title={\href{
https://doi.org/10.1155/2012/357802
}{Modified Chaplygin gas cosmology}},
  author={Benaoum, Hachemi B and others},
  journal={Advances in High Energy Physics},
  volume={2012},
  year={2012},
  publisher={Hindawi}
}

@article{guo2005observational,
  title={\href{https://doi.org/10.48550/arXiv.astro-ph/0509790}{Observational constraints on variable chaplygin gas}},
  author={Guo, Zong-Kuan and Zhang, Yuan-Zhong},
  journal={arXiv preprint astro-ph/0509790},
  year={2005}
}

@article{guo2007cosmology,
  title={\href{https://doi.org/10.48550/arXiv.astro-ph/0506091}{Cosmology with a variable Chaplygin gas}},
  author={Guo, Zong-Kuan and Zhang, Yuan-Zhong},
  journal={Physics Letters B},
  volume={645},
  number={4},
  pages={326--329},
  year={2007},
  publisher={Elsevier}
}

@article{chakraborty2019evolution,
  title={\href{https://doi.org/10.48550/arXiv.1906.12185}{Evolution of FRW universe in variable modified Chaplygin gas model}},
  author={Chakraborty, Samarjit and Guha, Sarbari and Panigrahi, D},
  journal={arXiv preprint arXiv:1906.12185},
  year={2019}
}

@article{villanueva2015jacobian,
  title={\href{http://dx.doi.org/10.1140/epjc/s10052-015-3464-z}{A Jacobian elliptic single-field inflation}},
  author={Villanueva, JR and Gallo, Emanuel},
  journal={The European Physical Journal C},
  volume={75},
  pages={1--7},
  year={2015},
  publisher={Springer}
}

@article{bahcall1999cosmic,
  title={\href{https://doi.org/10.1126/science.284.5419.1481}{The cosmic triangle: Revealing the state of the universe}},
  author={Bahcall, Neta A and Ostriker, Jeremiah P and Perlmutter, Saul and Steinhardt, Paul J},
  journal={Science},
  volume={284},
  number={5419},
  pages={1481--1488},
  year={1999},
  publisher={American Association for the Advancement of Science}
}

@article{perlmutter1999supernova,
  title={Supernova cosmology project collaboration},
  author={Perlmutter, S and others},
  journal={Astrophys. J},
  volume={517},
  number={2},
  pages={565},
  year={1999}
}

@article{filippenko1998results,
  title={\href{https://doi.org/10.1016/S0370-1573(98)00052-0}{Results from the high-z supernova search team}},
  author={Filippenko, Alexei V and Riess, Adam G},
  journal={Physics Reports},
  volume={307},
  number={1-4},
  pages={31--44},
  year={1998},
  publisher={Elsevier}
}

@article{allen2004constraints,
  title={\href{https://doi.org/10.1111/j.1365-2966.2004.08080.x}{Constraints on dark energy from Chandra observations of the largest relaxed galaxy clusters}},
  author={Allen, SW and Schmidt, Robert W and Ebeling, H and Fabian, AC and Van Speybroeck, L},
  journal={Monthly Notices of the Royal Astronomical Society},
  volume={353},
  number={2},
  pages={457--467},
  year={2004},
  publisher={Blackwell Science Ltd}
}

@article{abazajian2004second,
  title={\href{https://doi.org/10.1086/421365}{The second data release of the sloan digital sky survey}},
  author={Abazajian, Kevork and Adelman-McCarthy, Jennifer K and Ag{\"u}eros, Marcel A and Allam, Sahar S and Anderson, Kurt SJ and Anderson, Scott F and Annis, James and Bahcall, Neta A and Baldry, Ivan K and Bastian, Steven and others},
  journal={The Astronomical Journal},
  volume={128},
  number={1},
  pages={502},
  year={2004},
  publisher={IOP Publishing}
}

@article{perlmutter1999measurements,
  title={\href{https://dx.doi.org/10.1086/307221}{Measurements of $\Omega$ and $\Lambda$ from 42 high-redshift supernovae}},
  author={Perlmutter, Saul and Aldering, Goldhaber and Goldhaber, Gerson and Knop, RA and Nugent, Peter and Castro, Patricia G and Deustua, Susana and Fabbro, Sebastien and Goobar, Ariel and Groom, Donald E and others},
  journal={The Astrophysical Journal},
  volume={517},
  number={2},
  pages={565},
  year={1999},
  publisher={IOP Publishing}
}

@article{perlmutter1998discovery,
  title={\href{https://arxiv.org/abs/astro-ph/9712212}{Discovery of a supernova explosion at half the age of the Universe}},
  author={Perlmutter, Saul and Aldering, G and Valle, M Della and Deustua, S and Ellis, RS and Fabbro, S and Fruchter, A and Goldhaber, G and Groom, DE and Hook, IM and others},
  journal={Nature},
  volume={391},
  number={6662},
  pages={51--54},
  year={1998},
  publisher={Nature Publishing Group UK London}
}

@article{perlmutter1997measurements,
  title={\href{https://dx.doi.org/10.1086/304265}{Measurements* of the Cosmological Parameters $\Omega$ and $\Lambda$ from the First Seven Supernovae at z≥ 0.35}},
  author={Perlmutter, Saul and Gabi, S and Goldhaber, G and Goobar, A and Groom, DE and Hook, IM and Kim, AG and Kim, MY and Lee, JC and Pain, R and others},
  journal={The Astrophysical Journal},
  volume={483},
  number={2},
  pages={565},
  year={1997},
  publisher={IOP Publishing}
}

@article{tegmark2004cosmological,
  title={\href{https://doi.org/10.1103/PhysRevD.69.103501}{Cosmological parameters from SDSS and WMAP}},
  author={Tegmark, Max and Strauss, Michael A and Blanton, Michael R and Abazajian, Kevork and Dodelson, Scott and Sandvik, Havard and Wang, Xiaomin and Weinberg, David H and Zehavi, Idit and Bahcall, Neta A and others},
  journal={Physical Review D},
  volume={69},
  number={10},
  pages={103501},
  year={2004},
  publisher={APS}
}

@article{bennett2003first,
  title={First year Wilkinson Microwave Anisotropy Probe: Preliminary maps and basic results},
  author={Bennett, CL and others},
  journal={arXiv preprint astro-ph/0302207},
  year={2003}
}

@article{spergel2003first,
  title={\href{https://ui.adsabs.harvard.edu/link_gateway/2003ApJS..148..175S/doi:10.1086/377226}{First-year Wilkinson Microwave Anisotropy Probe (WMAP)* observations: determination of cosmological parameters}},
  author={Spergel, David N and Verde, Licia and Peiris, Hiranya V and Komatsu, Eiichiro and Nolta, MR and Bennett, Charles L and Halpern, Mark and Hinshaw, Gary and Jarosik, Norman and Kogut, Alan and others},
  journal={The Astrophysical Journal Supplement Series},
  volume={148},
  number={1},
  pages={175},
  year={2003},
  publisher={IOP Publishing}
}

@article{spergel2007three,
  title={\href{https://doi.org/10.1086/513700}{Three-year Wilkinson Microwave Anisotropy Probe (WMAP) observations: implications for cosmology}},
  author={Spergel, David N and Bean, R and Dor{\'e}, O and Nolta, MR and Bennett, CL and Dunkley, Joanna and Hinshaw, G and Jarosik, N ea and Komatsu, E and Page, L and others},
  journal={The Astrophysical Journal Supplement Series},
  volume={170},
  number={2},
  pages={377},
  year={2007},
  publisher={IOP Publishing}
}

@article{verde20022df,
  title={\href{https://doi.org/10.1046/j.1365-8711.2002.05620.x}{The 2dF Galaxy Redshift Survey: the bias of galaxies and the density of the Universe}},
  author={Verde, Licia and Heavens, Alan F and Percival, Will J and Matarrese, Sabino and Baugh, Carlton M and Bland-Hawthorn, Joss and Bridges, Terry and Cannon, Russell and Cole, Shaun and Colless, Matthew and others},
  journal={Monthly Notices of the Royal Astronomical Society},
  volume={335},
  number={2},
  pages={432--440},
  year={2002},
  publisher={The Royal Astronomical Society}
}

@article{eisenstein2005detection,
  title={\href{https://doi.org/10.1086/466512}{Detection of the baryon acoustic peak in the large-scale correlation function of SDSS luminous red galaxies}},
  author={Eisenstein, Daniel J and Zehavi, Idit and Hogg, David W and Scoccimarro, Roman and Blanton, Michael R and Nichol, Robert C and Scranton, Ryan and Seo, Hee-Jong and Tegmark, Max and Zheng, Zheng and others},
  journal={The Astrophysical Journal},
  volume={633},
  number={2},
  pages={560},
  year={2005},
  publisher={IOP Publishing}
}

@article{eisenstein1998baryonic,
  title={\href{https://doi.org/10.1086/305424}{Baryonic features in the matter transfer function}},
  author={Eisenstein, Daniel J and Hu, Wayne},
  journal={The Astrophysical Journal},
  volume={496},
  number={2},
  pages={605},
  year={1998},
  publisher={IOP Publishing}
}

@article{adam2016planck,
  title={\href{https://doi.org/10.1051/0004-6361/201527101}{Planck 2015 results-I. Overview of products and scientific results}},
  author={Adam, R{\'e}mi and Ade, Peter AR and Aghanim, N and Akrami, Y and Alves, MIR and Arg{\"u}eso, F and Arnaud, M and Arroja, F and Ashdown, Mark and Aumont, J and others},
  journal={Astronomy \& Astrophysics},
  volume={594},
  pages={A1},
  year={2016},
  publisher={EDP sciences}
}

@article{garnavich1998constraints,
  title={\href{https://doi.org/10.1086/311140}{Constraints on cosmological models from Hubble Space Telescope observations of high-z supernovae}},
  author={Garnavich, Peter M and Kirshner, Robert P and Challis, Peter and Tonry, John and Gilliland, Ron L and Smith, R Chris and Clocchiatti, Alejandro and Diercks, Alan and Filippenko, Alexei V and Hamuy, Mario and others},
  journal={The Astrophysical Journal},
  volume={493},
  number={2},
  pages={L53},
  year={1998},
  publisher={IOP Publishing}
}

@article{perlmutter1998cosmology,
  title={\href{https://arxiv.org/abs/astro-ph/9812473}{Cosmology from type Ia supernovae}},
  author={Perlmutter, Saul and Aldering, G and Deustua, S and Fabbro, S and Goldhaber, G and Groom, DE and Kim, AG and Kim, MY and Knop, RA and Nugent, P and others},
  journal={arXiv preprint astro-ph/9812473},
  year={1998}
}

@article{padmanabhan2003cosmological,
  title={\href{https://doi.org/10.1016/S0370-1573(03)00120-0}{Cosmological constant—the weight of the vacuum}},
  author={Padmanabhan, Thanu},
  journal={Physics reports},
  volume={380},
  number={5-6},
  pages={235--320},
  year={2003},
  publisher={Elsevier}
}

@article{carroll1992cosmological,
  title={\href{https://doi.org/10.1146/annurev.aa.30.090192.002435}{The cosmological constant}},
  author={Carroll, Sean M and Press, William H and Turner, Edwin L},
  journal={Annual Review of Astronomy and Astrophysics},
  volume={30},
  number={1},
  pages={499--542},
  year={1992},
  publisher={Annual Reviews 4139 El Camino Way, PO Box 10139, Palo Alto, CA 94303-0139, USA}
}

@article{peebles2003cosmological,
  title={\href{https://doi.org/10.1103/RevModPhys.75.559}{The cosmological constant and dark energy}},
  author={Peebles, P James E and Ratra, Bharat},
  journal={Reviews of Modern Physics},
  volume={75},
  number={2},
  pages={559},
  year={2003},
  publisher={APS}
}

@article{bianchi2010all,
  title={\href{https://doi.org/10.48550/arXiv.1002.3966}{Why all these prejudices against a constant?}},
  author={Bianchi, Eugenio and Rovelli, Carlo},
  journal={arXiv preprint arXiv:1002.3966},
  year={2010}
}

@article{weinberg2000cosmological,
  title={\href{https://doi.org/10.48550/arXiv.astro-ph/0005265}{The cosmological constant problems (talk given at Dark Matter 2000, february, 2000)}},
  author={Weinberg, Steven},
  journal={arXiv preprint astro-ph/0005265},
  year={2000}
}

@article{copeland2006dynamics,
  title={\href{https://doi.org/10.1142/S021827180600942X}{Dynamics of dark energy}},
  author={Copeland, Edmund J and Sami, Mohammad and Tsujikawa, Shinji},
  journal={International Journal of Modern Physics D},
  volume={15},
  number={11},
  pages={1753--1935},
  year={2006},
  publisher={World Scientific}
}

@article{armendariz2001essentials,
  title={\href{https://doi.org/10.1103/PhysRevD.63.103510}{Essentials of k-essence}},
  author={Armendariz-Picon, Christian and Mukhanov, V and Steinhardt, Paul J},
  journal={Physical Review D},
  volume={63},
  number={10},
  pages={103510},
  year={2001},
  publisher={APS}
}

@article{caldwell2002phantom,
  title={\href{https://doi.org/10.1016/S0370-2693(02)02589-3}{A phantom menace? Cosmological consequences of a dark energy component with super-negative equation of state}},
  author={Caldwell, Robert R},
  journal={Physics Letters B},
  volume={545},
  number={1-2},
  pages={23--29},
  year={2002},
  publisher={Elsevier}
}

@article{elizalde2004late,
  title={\href{https://doi.org/10.1103/PhysRevD.70.043539}{Late-time cosmology in a (phantom) scalar-tensor theory: dark energy and the cosmic speed-up}},
  author={Elizalde, Emilio and Nojiri, Shin’ichi and Odintsov, Sergei D},
  journal={Physical Review D},
  volume={70},
  number={4},
  pages={043539},
  year={2004},
  publisher={APS}
}

@article{caldwell1998cosmological,
  title={\href{https://doi.org/10.1103/PhysRevLett.80.1582}{Cosmological imprint of an energy component with general equation of state}},
  author={Caldwell, Robert R and Dave, Rahul and Steinhardt, Paul J},
  journal={Physical Review Letters},
  volume={80},
  number={8},
  pages={1582},
  year={1998},
  publisher={APS}
}

@article{sen2002rolling,
  title={\href{https://doi.org/10.1088/1126-6708/2002/04/048}{Rolling tachyon}},
  author={Sen, Ashoke},
  journal={Journal of High Energy Physics},
  volume={2002},
  number={04},
  pages={048},
  year={2002},
  publisher={IOP Publishing}
}

@article{gibbons2002cosmological,
  title={\href{https://doi.org/10.1016/S0370-2693(02)01881-6}{Cosmological evolution of the rolling tachyon}},
  author={Gibbons, GW},
  journal={Physics Letters B},
  volume={537},
  number={1-2},
  pages={1--4},
  year={2002},
  publisher={Elsevier}
}

@article{martin2008dbi,
  title={\href{https://doi.org/10.1103/PhysRevD.77.123508}{DBI-essence}},
  author={Martin, Jerome and Yamaguchi, Masahide},
  journal={Physical Review D},
  volume={77},
  number={12},
  pages={123508},
  year={2008},
  publisher={APS}
}

@article{chakraborty2020investigating,
  title={\href{https://doi.org/10.1142/S021827182050087X}{Investigating inflation driven by DBI-essence scalar field}},
  author={Chakraborty, Gargee and Chattopadhyay, Surajit},
  journal={International Journal of Modern Physics D},
  volume={29},
  number={12},
  pages={2050087},
  year={2020},
  publisher={World Scientific}
}

@article{peebles1988cosmology,
  title={\href{https://doi.org/10.1086/185100}{Cosmology with a time-variable cosmological'constant'}},
  author={Peebles, PJE and Ratra, Bharat},
  journal={Astrophysical Journal, Part 2-Letters to the Editor (ISSN 0004-637X), vol. 325, Feb. 15, 1988, p. L17-L20. NSF-supported research.},
  volume={325},
  pages={L17--L20},
  year={1988}
}

@article{bamba2012dark,
  title={\href{https://doi.org/10.1007/s10509-012-1181-8}{Dark energy cosmology: the equivalent description via different theoretical models and cosmography tests}},
  author={Bamba, Kazuharu and Capozziello, Salvatore and Nojiri, Shin’ichi and Odintsov, Sergei D},
  journal={Astrophysics and Space Science},
  volume={342},
  pages={155--228},
  year={2012},
  publisher={Springer}
}

@article{chen2015constraints,
  title={\href{https://ui.adsabs.harvard.edu/link_gateway/2015JCAP...02..010C/doi:10.1088/1475-7516/2015/02/010}{Constraints on a $\phi$CDM model from strong gravitational lensing and updated Hubble parameter measurements}},
  author={Chen, Yun and Geng, Chao-Qiang and Cao, Shuo and Huang, Yu-Mei and Zhu, Zong-Hong},
  journal={Journal of Cosmology and Astroparticle Physics},
  volume={2015},
  number={02},
  pages={010},
  year={2015},
  publisher={IOP Publishing}
}

@article{smer2017planck,
  title={\href{https://doi.org/10.1088/1475-7516/2017/01/023}{Planck satellite constraints on pseudo-Nambu-Goldstone boson quintessence}},
  author={Smer-Barreto, Vanessa and Liddle, Andrew R},
  journal={Journal of Cosmology and Astroparticle Physics},
  volume={2017},
  number={01},
  pages={023},
  year={2017},
  publisher={IOP Publishing}
}

@article{wetterich1988cosmology,
  title={\href{https://doi.org/10.1016/0550-3213(88)90193-9}{Cosmology and the fate of dilatation symmetry}},
  author={Wetterich, Christof},
  journal={Nuclear Physics B},
  volume={302},
  number={4},
  pages={668--696},
  year={1988},
  publisher={Elsevier}
}

@article{weinberg1989cosmological,
  title={\href{https://doi.org/10.1103/RevModPhys.61.1}{The cosmological constant problem}},
  author={Weinberg, Steven},
  journal={Reviews of Modern Physics},
  volume={61},
  number={1},
  pages={1},
  year={1989},
  publisher={APS}
}

@article{armendariz2000dynamical,
  title={\href{https://doi.org/10.1103/PhysRevLett.85.4438}{Dynamical solution to the problem of a small cosmological constant and late-time cosmic acceleration}},
  author={Armendariz-Picon, Christian and Mukhanov, V and Steinhardt, Paul J},
  journal={Physical Review Letters},
  volume={85},
  number={21},
  pages={4438},
  year={2000},
  publisher={APS}
}

@article{sen2002tachyon,
  title={\href{https://doi.org/10.1088/1126-6708/2002/07/065}{Tachyon matter}},
  author={Sen, Ashoke},
  journal={Journal of High Energy Physics},
  volume={2002},
  number={07},
  pages={065},
  year={2002},
  publisher={IOP Publishing}
}

@article{feng2005dark,
  title={\href{https://doi.org/10.1016/j.physletb.2004.12.071}{Dark energy constraints from the cosmic age and supernova}},
  author={Feng, Bo and Wang, Xiulian and Zhang, Xinmin},
  journal={Physics Letters B},
  volume={607},
  number={1-2},
  pages={35--41},
  year={2005},
  publisher={Elsevier}
}

@article{guo2005cosmological,
  title={\href{https://doi.org/10.1016/j.physletb.2005.01.017}{Cosmological evolution of a quintom model of dark energy}},
  author={Guo, Zong-Kuan and Piao, Yun-Song and Zhang, Xinmin and Zhang, Yuan-Zhong},
  journal={Physics Letters B},
  volume={608},
  number={3-4},
  pages={177--182},
  year={2005},
  publisher={Elsevier}
}

@article{gasperini2001quintessence,
  title={\href{https://doi.org/10.1103/PhysRevD.65.023508}{Quintessence as a runaway dilaton}},
  author={Gasperini, Maurizio and Piazza, F and Veneziano, G},
  journal={Physical Review D},
  volume={65},
  number={2},
  pages={023508},
  year={2001},
  publisher={APS}
}

@article{gumjudpai2009generalized,
  title={\href{https://doi.org/10.1103/PhysRevD.80.023528}{Generalized DBI quintessence}},
  author={Gumjudpai, Burin and Ward, John},
  journal={Physical Review D},
  volume={80},
  number={2},
  pages={023528},
  year={2009},
  publisher={APS}
}

@article{wei2005hessence,
  title={\href{https://doi.org/10.1088/0264-9381/22/16/005}{Hessence: a new view of quintom dark energy}},
  author={Wei, Hao and Cai, Rong-Gen and Zeng, Ding-Fang},
  journal={Classical and Quantum Gravity},
  volume={22},
  number={16},
  pages={3189},
  year={2005},
  publisher={IOP Publishing}
}

@article{dvali20004d,
  title={\href{https://doi.org/10.1016/S0370-2693(00)00669-9}{4D gravity on a brane in 5D Minkowski space}},
  author={Dvali, Gia and Gabadadze, Gregory and Porrati, Massimo},
  journal={Physics Letters B},
  volume={485},
  number={1-3},
  pages={208--214},
  year={2000},
  publisher={Elsevier}
}

@article{jacobson2001gravity,
  title={\href{https://doi.org/10.1103/PhysRevD.64.024028}{Gravity with a dynamical preferred frame}},
  author={Jacobson, Ted and Mattingly, David},
  journal={Physical Review D},
  volume={64},
  number={2},
  pages={024028},
  year={2001},
  publisher={APS}
}

@article{abdalla2005consistent,
  title={\href{https://doi.org/10.1088/0264-9381/22/5/L01}{Consistent modified gravity: dark energy, acceleration and the absence of cosmic doomsday}},
  author={Abdalla, Maria Christina B and Nojiri, Shin'ichi and Odintsov, Sergei D},
  journal={Classical and Quantum Gravity},
  volume={22},
  number={5},
  pages={L35},
  year={2005},
  publisher={IOP Publishing}
}

@article{hovrava2009membranes,
  title={\href{https://doi.org/10.1088/1126-6708/2009/03/020}{Membranes at quantum criticality}},
  author={Ho{\v{r}}ava, Petr},
  journal={Journal of High Energy Physics},
  volume={2009},
  number={03},
  pages={020},
  year={2009},
  publisher={IOP Publishing}
}

@article{barrow2012some,
  title={\href{https://doi.org/10.1103/PhysRevD.85.047503}{Some inflationary Einstein-aether cosmologies}},
  author={Barrow, John D},
  journal={Physical Review D},
  volume={85},
  number={4},
  pages={047503},
  year={2012},
  publisher={APS}
}

@article{meng2012einstein,
  title={\href{https://doi.org/10.1016/j.physletb.2012.03.024}{Einstein-aether theory as an alternative to dark energy model?}},
  author={Meng, Xinhe and Du, Xiaolong},
  journal={Physics Letters B},
  volume={710},
  number={4-5},
  pages={493--499},
  year={2012},
  publisher={Elsevier}
}

@article{nojiri2005modified,
  title={\href{https://doi.org/10.1016/j.physletb.2005.10.010}{Modified Gauss--Bonnet theory as gravitational alternative for dark energy}},
  author={Nojiri, Shin'ichi and Odintsov, Sergei D},
  journal={Physics Letters B},
  volume={631},
  number={1-2},
  pages={1--6},
  year={2005},
  publisher={Elsevier}
}

@incollection{tsujikawa2010modified,
  title={\href{https://doi.org/10.1007/978-3-642-10598-2_3}{Modified gravity models of dark energy}},
  author={Tsujikawa, Shinji},
  booktitle={Lectures on Cosmology: Accelerated Expansion of the Universe},
  pages={99--145},
  year={2010},
  publisher={Springer}
}

@article{bilic2002unification,
  title={\href{https://doi.org/10.1016/S0370-2693(02)01716-1}{Unification of dark matter and dark energy: the inhomogeneous Chaplygin gas}},
  author={Bili{\'c}, Neven and Tupper, Gary B and Viollier, Raoul D},
  journal={Physics Letters B},
  volume={535},
  number={1-4},
  pages={17--21},
  year={2002},
  publisher={Elsevier}
}

@inproceedings{gorini2005chaplygin,
  title={\href{https://doi.org/10.1142/9789812704030_0050}{The Chaplygin gas as a model for dark energy}},
  author={Gorini, Vittorio and Kamenshchik, Alexander and Moschella, Ugo and Pasquier, Vincent},
  booktitle={The Tenth Marcel Grossmann Meeting: On Recent Developments in Theoretical and Experimental General Relativity, Gravitation and Relativistic Field Theories (In 3 Volumes)},
  pages={840--859},
  year={2005},
  organization={World Scientific}
}

@article{alam2003exploring,
  title={\href{https://doi.org/10.1046/j.1365-8711.2003.06871.x}{Exploring the expanding universe and dark energy using the Statefinder diagnostic}},
  author={Alam, Ujjaini and Sahni, Varun and Deep Saini, Tarun and Starobinsky, AA},
  journal={Monthly Notices of the Royal Astronomical Society},
  volume={344},
  number={4},
  pages={1057--1074},
  year={2003},
  publisher={Blackwell Science Ltd Oxford, UK}
}

@article{bento2002generalized,
  title={\href{https://doi.org/10.1103/PhysRevD.66.043507}{Generalized Chaplygin gas, accelerated expansion, and dark-energy-matter unification}},
  author={Bento, MC and Bertolami, Orfeu and Sen, Anjan A},
  journal={Physical Review D},
  volume={66},
  number={4},
  pages={043507},
  year={2002},
  publisher={APS}
}

@article{makler2003constraints,
  title={\href{https://doi.org/10.1016/S0370-2693(03)00038-8}{Constraints on the generalized Chaplygin gas from supernovae observations}},
  author={Makler, Mart{\i}n and de Oliveira, S{\'e}rgio Quinet and Waga, Ioav},
  journal={Physics Letters B},
  volume={555},
  number={1-2},
  pages={1--6},
  year={2003},
  publisher={Elsevier}
}

@article{zhai2006viscous,
  title={\href{https://doi.org/10.1142/S0218271806008784}{Viscous generalized Chaplygin gas}},
  author={Zhai, Xiang-Hua and Xu, You-Dong and Li, Xin-Zhou},
  journal={International Journal of Modern Physics D},
  volume={15},
  number={08},
  pages={1151--1161},
  year={2006},
  publisher={World Scientific}
}

@article{bhadra2012accretion,
  title={\href{https://doi.org/10.1140/epjc/s10052-012-1912-6}{Accretion of new variable modified Chaplygin gas and generalized cosmic Chaplygin gas onto Schwarzschild and Kerr--Newman black holes}},
  author={Bhadra, Jhumpa and Debnath, Ujjal},
  journal={The European Physical Journal C},
  volume={72},
  pages={1--9},
  year={2012},
  publisher={Springer}
}

@article{sharif2014effects,
  title={\href{https://doi.org/10.1088/1475-7516/2014/12/038}{Effects of viscous pressure on warm inflationary generalized cosmic Chaplygin gas model}},
  author={Sharif, M and Saleem, Rabia},
  journal={Journal of Cosmology and Astroparticle Physics},
  volume={2014},
  number={12},
  pages={038},
  year={2014},
  publisher={IOP Publishing}
}

@article{rudra2012dynamics,
  title={\href{https://doi.org/10.1007/s10509-012-1198-z}{Dynamics of interacting generalized cosmic Chaplygin gas in brane-world scenario}},
  author={Rudra, Prabir},
  journal={Astrophysics and Space Science},
  volume={342},
  pages={579--599},
  year={2012},
  publisher={Springer}
}

@article{sharif2014phantom,
  title={\href{https://doi.org/10.1140/epjp/i2014-14015-5}{Phantom-like generalized cosmic chaplygin gas and traversable wormhole solutions}},
  author={Sharif, M and Jawad, Abdul},
  journal={The European Physical Journal Plus},
  volume={129},
  pages={1--13},
  year={2014},
  publisher={Springer}
}

@article{eid2018schwarzschild,
  title={\href{https://ui.adsabs.harvard.edu/link_gateway/2018GrCo...24..378E/doi:10.1134/S0202289318040072}{Schwarzschild--De Sitter thin shell wormholes supported by a generalized cosmic Chaplygin gas}},
  author={Eid, Ali},
  journal={Gravitation and Cosmology},
  volume={24},
  number={4},
  pages={378--383},
  year={2018},
  publisher={Springer}
}

@article{sharif2013reissner,
  title={\href{https://doi.org/10.1140/EPJC/S10052-013-2554-Z}{Reissner--Nordstr{\"o}m thin-shell wormholes with generalized cosmic Chaplygin gas}},
  author={Sharif, M and Azam, M},
  journal={The European Physical Journal C},
  volume={73},
  pages={1--6},
  year={2013},
  publisher={Springer}
}

@article{panigrahi2016thermodynamics,
  title={\href{https://doi.org/10.1088/1475-7516/2016/05/052}{Thermodynamics of the variable modified Chaplygin gas}},
  author={Panigrahi, D and Chatterjee, S},
  journal={Journal of Cosmology and Astroparticle Physics},
  volume={2016},
  number={05},
  pages={052},
  year={2016},
  publisher={IOP Publishing}
}

@article{jamil2008interacting,
  title={\href{http://dx.doi.org/10.1140/epjc/s10052-008-0722-3}{Interacting modified variable Chaplygin gas in a non-flat universe}},
  author={Jamil, Mubasher and Rashid, Muneer Ahmad},
  journal={The European Physical Journal C},
  volume={58},
  pages={111--114},
  year={2008},
  publisher={Springer}
}

@article{chattopadhyay2009holographic,
  title={\href{https://doi.org/10.1007/s10509-009-9977-x}{Holographic dark energy scenario and variable modified Chaplygin gas}},
  author={Chattopadhyay, Surajit and Debnath, Ujjal},
  journal={Astrophysics and Space Science},
  volume={319},
  pages={183--185},
  year={2009},
  publisher={Springer}
}

@article{rudra2013effective,
  title={\href{https://doi.org/10.1007/s10773-012-1150-6}{How effective is new variable modified Chaplygin gas to play the role of dark energy—a dynamical system analysis in RS II brane model}},
  author={Rudra, Prabir and Ranjit, Chayan and Kundu, Sujata},
  journal={Astrophysics and Space Science},
  volume={347},
  pages={433--444},
  year={2013},
  publisher={Springer}
}

@article{mukherjee2023accretion,
  title={\href{https://doi.org/10.1142/S0219887823502183}{Accretion of Modified Chaplygin-Jacobi Gas and Modified Chaplygin-Abel Gas onto Schwarzschild Black Hole}},
  author={Mukherjee, Puja and Debnath, Ujjal and Pradhan, Anirudh},
  journal={International Journal of Geometric Methods in Modern Physics},
  year={2023},
  publisher={World Scientific}
}

@article{chaudhary2023constraints,
  title={\href{https://doi.org/10.48550/arXiv.2307.14691}{Constraints on the Parameters of Modified Chaplygin-Jacobi and Modified Chaplygin-Abel Gases in $ f (T) $ Gravity Model}},
  author={Chaudhary, Himanshu and Debnath, Ujjal and Roy, Tanusree and Maity, Sayani and Mustafa, G},
  journal={arXiv preprint arXiv:2307.14691},
  year={2023}
}

@article{michel1972accretion,
  title={\href{https://ui.adsabs.harvard.edu/link_gateway/1972Ap&SS..15..153M/doi:10.1007/BF00649949}{Accretion of matter by condensed objects}},
  author={Michel, F Curtis},
  journal={Astrophysics and Space Science},
  volume={15},
  pages={153--160},
  year={1972},
  publisher={Springer}
}

@article{babichev2004black,
  title={\href{https://doi.org/10.1103/PhysRevLett.93.021102}{Black hole mass decreasing due to phantom energy accretion}},
  author={Babichev, Eugeny and Dokuchaev, Vyacheslav and Eroshenko, Yu},
  journal={Physical Review Letters},
  volume={93},
  number={2},
  pages={021102},
  year={2004},
  publisher={APS}
}

@article{babichev2005accretion,
  title={\href{https://ui.adsabs.harvard.edu/link_gateway/2005JETP..100..528B/doi:10.1134/1.1901765}{The accretion of dark energy onto a black hole}},
  author={Babichev, EO and Dokuchaev, VI and Eroshenko, Yu N},
  journal={Journal of Experimental and Theoretical Physics},
  volume={100},
  pages={528--538},
  year={2005},
  publisher={Springer}
}

@article{jamil2008charged,
  title={\href{https://doi.org/10.1140/EPJC/S10052-008-0761-9}{Charged black holes in phantom cosmology}},
  author={Jamil, Mubasher and Qadir, Asghar and Rashid, Muneer Ahmad},
  journal={The European Physical Journal C},
  volume={58},
  pages={325--329},
  year={2008},
  publisher={Springer}
}

@article{jimenez2008evolution,
  title={\href{https://ui.adsabs.harvard.edu/link_gateway/2008GrCo...14..213J/doi:10.1134/S020228930803002X}{Evolution of a Kerr-Newman black hole in a dark energy universe}},
  author={Jim{\'e}nez Madrid, Jose A and Gonzalez-D{\'\i}az, Pedro F},
  journal={Gravitation and Cosmology},
  volume={14},
  pages={213--225},
  year={2008},
  publisher={Springer}
}

@article{sharif2012phantom,
  title={\href{https://doi.org/10.1088/0256-307X/29/1/010401}{Phantom energy accretion by a stringy charged black hole}},
  author={Sharif, M and Abbas, G},
  journal={Chinese Physics Letters},
  volume={29},
  number={1},
  pages={010401},
  year={2012},
  publisher={IOP Publishing}
}

@article{sun2008phantom,
  title={\href{https://doi.org/10.1103/PhysRevD.78.064060}{Phantom energy accretion onto black holes in a cyclic universe}},
  author={Sun, Cheng-Yi},
  journal={Physical Review D},
  volume={78},
  number={6},
  pages={064060},
  year={2008},
  publisher={APS}
}

@article{abbas2013thermodynamics,
  title={\href{https://doi.org/10.1088/0256-307X/30/10/100403}{Thermodynamics of Phantom Energy Accreting onto a Black Hole in Einstein—Power—Maxwell Gravity}},
  author={Abbas, G and Ramzan, RM},
  journal={Chinese Physics Letters},
  volume={30},
  number={10},
  pages={100403},
  year={2013},
  publisher={IOP Publishing}
}

@article{debnath2015accretion,
  title={\href{https://doi.org/10.1140/epjc/s10052-015-3349-1}{Accretion and evaporation of modified Hayward black hole}},
  author={Debnath, Ujjal},
  journal={The European Physical Journal C},
  volume={75},
  number={3},
  pages={129},
  year={2015},
  publisher={Springer}
}

@article{hendi2011charged,
  title={\href{http://dx.doi.org/10.1140/epjc/s10052-011-1551-3}{Charged BTZ-like black holes in higher dimensions}},
  author={Hendi, SH},
  journal={The European Physical Journal C},
  volume={71},
  number={2},
  pages={1551},
  year={2011},
  publisher={Springer}
}

@article{kim1997renormalized,
  title={\href{https://doi.org/10.1103/PhysRevD.55.2159}{Renormalized thermodynamic entropy of black holes in higher dimensions}},
  author={Kim, Sang Pyo and Kim, Sung Ku and Soh, Kwang-Sup and Yee, Jae Hyung},
  journal={Physical Review D},
  volume={55},
  number={4},
  pages={2159},
  year={1997},
  publisher={APS}
}

@article{hovrava2009spectral,
  title={\href{https://doi.org/10.1103/PhysRevLett.102.161301}{Spectral dimension of the universe in quantum gravity at a Lifshitz point}},
  author={Ho{\v{r}}ava, Petr},
  journal={Physical Review Letters},
  volume={102},
  number={16},
  pages={161301},
  year={2009},
  publisher={APS}
}

@article{cai2009dynamical,
  title={\href{https://doi.org/10.1103/PhysRevD.80.041501}{Dynamical scalar degree of freedom in Ho{\v{r}}ava-Lifshitz gravity}},
  author={Cai, Rong-Gen and Hu, Bin and Zhang, Hong-Bo},
  journal={Physical Review D},
  volume={80},
  number={4},
  pages={041501},
  year={2009},
  publisher={APS}
}

@article{calcagni2009cosmology,
  title={\href{https://doi.org/10.1088/1126-6708/2009/09/112}{Cosmology of the Lifshitz universe}},
  author={Calcagni, Gianluca},
  journal={Journal of High Energy Physics},
  volume={2009},
  number={09},
  pages={112},
  year={2009},
  publisher={IOP Publishing}
}

@article{kiritsis2009hovrava,
  title={\href{https://doi.org/10.1016/j.nuclphysb.2009.05.005}{Ho{\v{r}}ava--lifshitz cosmology}},
  author={Kiritsis, Elias and Kofinas, Georgios},
  journal={Nuclear Physics B},
  volume={821},
  number={3},
  pages={467--480},
  year={2009},
  publisher={Elsevier}
}

@article{takahashi2009chiral,
  title={\href{https://doi.org/10.1103/PhysRevLett.102.231301}{Chiral primordial gravitational waves from a Lifshitz point}},
  author={Takahashi, Tomohiro and Soda, Jiro},
  journal={Physical Review Letters},
  volume={102},
  number={23},
  pages={231301},
  year={2009},
  publisher={APS}
}

@article{mukohyama2009scale,
  title={\href{https://doi.org/10.1088/1475-7516/2009/06/001}{Scale-invariant cosmological perturbations from Ho{\v{r}}ava-Lifshitz gravity without inflation}},
  author={Mukohyama, Shinji},
  journal={Journal of Cosmology and Astroparticle Physics},
  volume={2009},
  number={06},
  pages={001},
  year={2009},
  publisher={IOP Publishing}
}

@article{mukohyama2009phenomenological,
  title={\href{https://doi.org/10.1016/j.physletb.2009.07.005}{Phenomenological aspects of Ho{\v{r}}ava--Lifshitz cosmology}},
  author={Mukohyama, Shinji and Nakayama, Kazunori and Takahashi, Fuminobu and Yokoyama, Shuichiro},
  journal={Physics Letters B},
  volume={679},
  number={1},
  pages={6--9},
  year={2009},
  publisher={Elsevier}
}

@article{lu2009solutions,
  title={\href{https://doi.org/10.1103/PhysRevLett.103.091301}{Solutions to Ho{\v{r}}ava gravity}},
  author={L{\"u}, H and Mei, Jianwei and Pope, CN},
  journal={Physical Review Letters},
  volume={103},
  number={9},
  pages={091301},
  year={2009},
  publisher={APS}
}

@article{leon2009phase,
  title={\href{https://doi.org/10.1088/1475-7516/2009/11/006}{Phase-space analysis of Ho{\v{r}}ava-Lifshitz cosmology}},
  author={Leon, Genly and Saridakis, Emmanuel N},
  journal={Journal of Cosmology and Astroparticle Physics},
  volume={2009},
  number={11},
  pages={006},
  year={2009},
  publisher={IOP Publishing}
}

@article{minamitsuji2010classification,
  title={\href{https://doi.org/10.1016/j.physletb.2010.01.021}{Classification of cosmology with arbitrary matter in the Ho{\v{r}}ava--Lifshitz model}},
  author={Minamitsuji, Masato},
  journal={Physics Letters B},
  volume={684},
  number={4-5},
  pages={194--198},
  year={2010},
  publisher={Elsevier}
}

@article{park2010test,
  title={\href{http://dx.doi.org/10.1088/1475-7516/2010/01/001}{A test of Ho{\v{r}}ava gravity: the dark energy}},
  author={Park, Mu-In},
  journal={Journal of Cosmology and Astroparticle Physics},
  volume={2010},
  number={01},
  pages={001},
  year={2010},
  publisher={IOP Publishing}
}

@article{setare2010holographic,
  title={\href{https://doi.org/10.1088/1475-7516/2010/02/010}{Holographic dark energy with varying gravitational constant in Ho{\v{r}}ava-Lifshitz cosmology}},
  author={Setare, MR and Jamil, Mubasher},
  journal={Journal of Cosmology and Astroparticle Physics},
  volume={2010},
  number={02},
  pages={010},
  year={2010},
  publisher={IOP Publishing}
}

@article{kim2009surplus,
  title={\href{https://doi.org/10.1103/PhysRevD.80.124002}{Surplus solid angle as an imprint of Ho{\v{r}}ava-Lifshitz gravity}},
  author={Kim, Sung-Soo and Kim, Taekyung and Kim, Yoonbai},
  journal={Physical Review D},
  volume={80},
  number={12},
  pages={124002},
  year={2009},
  publisher={APS}
}

@article{izumi2010stellar,
  title={\href{https://doi.org/10.1103/PhysRevD.81.044008}{Stellar center is dynamical in Ho{\v{r}}ava-Lifshitz gravity}},
  author={Izumi, Keisuke and Mukohyama, Shinji},
  journal={Physical Review D},
  volume={81},
  number={4},
  pages={044008},
  year={2010},
  publisher={APS}
}

@article{dutta2009observational,
  title={\href{https://doi.org/10.1088/1475-7516/2010/01/013}{Observational constraints on Horava-Lifshitz cosmology}},
  author={Dutta, Sourish and Saridakis, Emmanuel N},
  journal={arXiv preprint arXiv:0911.1435},
  year={2009}
}

@article{kehagias2009black,
  title={\href{https://doi.org/10.1016/j.physletb.2009.06.019}{The black hole and FRW geometries of non-relativistic gravity}},
  author={Kehagias, Alex and Sfetsos, Konstadinos},
  journal={Physics Letters B},
  volume={678},
  number={1},
  pages={123--126},
  year={2009},
  publisher={Elsevier}
}

@article{cai2009topological,
  title={\href{https://doi.org/10.1103/PhysRevD.80.024003}{Topological black holes in Ho{\v{r}}ava-Lifshitz gravity}},
  author={Cai, Rong-Gen and Cao, Li-Ming and Ohta, Nobuyoshi},
  journal={Physical Review D},
  volume={80},
  number={2},
  pages={024003},
  year={2009},
  publisher={APS}
}

@article{myung2009thermodynamics,
  title={\href{https://doi.org/10.1016/j.physletb.2009.06.013}{Thermodynamics of black holes in the deformed Ho{\v{r}}ava--Lifshitz gravity}},
  author={Myung, Yun Soo},
  journal={Physics Letters B},
  volume={678},
  number={1},
  pages={127--130},
  year={2009},
  publisher={Elsevier}
}

@article{park2009black,
  title={\href{http://dx.doi.org/10.1088/1126-6708/2009/09/123}{The black hole and cosmological solutions in IR modified Ho{\v{r}}ava gravity}},
  author={Park, Mu-in},
  journal={Journal of High Energy Physics},
  volume={2009},
  number={09},
  pages={123},
  year={2009},
  publisher={IOP Publishing}
}

@article{cai2010horizon,
  title={\href{https://doi.org/10.1103/PhysRevD.81.084061}{Horizon thermodynamics and gravitational field equations in Ho{\v{r}}ava-Lifshitz gravity}},
  author={Cai, Rong-Gen and Ohta, Nobuyoshi},
  journal={Physical Review D},
  volume={81},
  number={8},
  pages={084061},
  year={2010},
  publisher={APS}
}

@article{wei2019geodesics,
  title={\href{https://doi.org/10.1103/PhysRevD.99.104016}{Geodesics and periodic orbits in Kehagias-Sfetsos black holes in deformed Horava-Lifshitz gravity}},
  author={Wei, Shao-Wen and Yang, Jie and Liu, Yu-Xiao},
  journal={Physical Review D},
  volume={99},
  number={10},
  pages={104016},
  year={2019},
  publisher={APS}
}

@article{cai2009thermodynamics,
  title={\href{https://doi.org/10.1016/j.physletb.2009.07.075}{Thermodynamics of black holes in Ho{\v{r}}ava--Lifshitz gravity}},
  author={Cai, Rong-Gen and Cao, Li-Ming and Ohta, Nobuyoshi},
  journal={Physics Letters B},
  volume={679},
  number={5},
  pages={504--509},
  year={2009},
  publisher={Elsevier}
}

@article{arnowitt1961coordinate,
  title={\href{https://doi.org/10.1103/PhysRev.122.997}{Coordinate invariance and energy expressions in general relativity}},
  author={Arnowitt, Richard and Deser, Stanley and Misner, Charles W},
  journal={Physical Review},
  volume={122},
  number={3},
  pages={997},
  year={1961},
  publisher={APS}
}

@article{sotiriou2009quantum,
  title={\href{https://doi.org/10.1088/1126-6708/2009/10/033}{Quantum gravity without Lorentz invariance}},
  author={Sotiriou, Thomas P and Visser, Matt and Weinfurtner, Silke},
  journal={Journal of High Energy Physics},
  volume={2009},
  number={10},
  pages={033},
  year={2009},
  publisher={IOP Publishing}
}

@article{culetu2015source,
  title={\href{https://ui.adsabs.harvard.edu/link_gateway/2015Ap&SS.360...36C/doi:10.1007/s10509-015-2549-3}{On the source of the Kehagias-Sfetsos black hole}},
  author={Culetu, Hristu},
  journal={Astrophysics and Space Science},
  volume={360},
  pages={1--3},
  year={2015},
  publisher={Springer}
}

@article{debnath2021observational,
  title={\href{https://doi.org/10.1142/S0217751X21501578}{Observational data analysis for generalized cosmic Chaplygin gas in the background of Brans--Dicke theory}},
  author={Debnath, Ujjal},
  journal={International Journal of Modern Physics A},
  volume={36},
  number={21},
  pages={2150157},
  year={2021},
  publisher={World Scientific}
}

@article{debnath2020gravitational,
  title={\href{https://doi.org/10.1140/epjp/s13360-020-00219-9}{Gravitational waves for variable modified Chaplygin gas and some parametrizations of dark energy in the background of FRW universe}},
  author={Debnath, Ujjal},
  journal={The European Physical Journal Plus},
  volume={135},
  pages={1--22},
  year={2020},
  publisher={Springer}
}

@article{debnath2014constraining,
  title={\href{https://doi.org/10.1155/2014/653630}{Constraining the parameters of modified chaplygin gas in Einstein-Aether gravity}},
  author={Debnath, Ujjal and others},
  journal={Advances in High Energy Physics},
  volume={2014},
  year={2014},
  publisher={Hindawi}
}

@article{debnath2021constraining,
  title={\href{https://doi.org/10.1016/j.dark.2020.100764}{Constraining the parameters of modified Chaplygin gas in Brans--Dicke theory}},
  author={Debnath, Ujjal},
  journal={Physics of the Dark Universe},
  volume={31},
  pages={100764},
  year={2021},
  publisher={Elsevier}
}

@article{debnath2021gravitational,
  title={\href{https://doi.org/10.1016/j.dark.2021.100832}{Gravitational waves for some dark energy models in FRW Universe}},
  author={Debnath, Ujjal},
  journal={Physics of the Dark Universe},
  volume={32},
  pages={100832},
  year={2021},
  publisher={Elsevier}
}

@article{aberkane2017viscous,
  title={\href{https://doi.org/10.1088/0256-307X/34/6/069801}{Viscous modified Chaplygin gas in classical and loop quantum cosmology}},
  author={Aberkane, D and Mebarki, N and Benchikh, S},
  journal={Chinese Physics Letters},
  volume={34},
  number={6},
  pages={069801},
  year={2017},
  publisher={IOP Publishing}
}

@article{ranjit2016observational,
  title={Observational data fitting to constrain Variable Modified Chaplygin Gas in the background of Horava-Lifshitz Gravity},
  author={Ranjit, Chayan and Rudra, Prabir},
  journal={International Journal of Theoretical Physics},
  volume={55},
  pages={636--647},
  year={2016},
  publisher={Springer}
}

@article{geng2015accretion,
  title={\href{https://doi.org/10.1007/s10509-015-2519-9}{Accretion of dark energy onto stringy electrically charged black hole}},
  author={Geng, Jin-Ling and Zhang, Yu and Li, En-Kun and Duan, Peng-Fei},
  journal={Astrophysics and Space Science},
  volume={360},
  number={1},
  pages={1--7},
  year={2015},
  publisher={Springer}
}

@article{babichev2013black,
  title={\href{https://doi.org/10.3367/UFNe.0183.201312a.1257}{Black holes in the presence of dark energy}},
  author={Babichev, Eugeniy Olegovich and Dokuchaev, Vyacheslav Ivanovich and Eroshenko, Yu N},
  journal={Physics-Uspekhi},
  volume={56},
  number={12},
  pages={1155},
  year={2013},
  publisher={IOP Publishing}
}

@article{mukherjee2024constraining,
  title={\href{https://doi.org/10.1140/epjc/s10052-024-13196-5}{Constraining the parameters of generalized and viscous modified Chaplygin gas and black hole accretion in Einstein-Aether gravity}},
  author={Mukherjee, Puja and Debnath, Ujjal and Chaudhary, Himanshu and Mustafa, G},
  journal={The European Physical Journal C},
  volume={84},
  number={9},
  pages={930},
  year={2024},
  publisher={Springer}
}

@article{paramanik2024non,
  title={\href{https://ui.adsabs.harvard.edu/link_gateway/2024IJGMM..2150154P/doi:10.1142/S0219887824501548}{Non-commutative Wormhole geometries in presence of modified Chaplygin-Jacobi gas and Anton-Schmidt fluid}},
  author={Paramanik, Soubhik and Debnath, Ujjal},
  journal={International Journal of Geometric Methods in Modern Physics},
  volume={21},
  number={8},
  pages={2450154--847},
  year={2024}
}

@article{jamil2011accretion,
  title={\href{https://doi.org/10.1007/s10773-010-0553-5}{Accretion of phantom energy and generalized second law of thermodynamics for Einstein-Maxwell-Gauss-Bonnet black hole}},
  author={Jamil, Mubasher and Hussain, Ibrar},
  journal={International Journal of Theoretical Physics},
  volume={50},
  pages={465--472},
  year={2011},
  publisher={Springer}
}

@article{jamil2011generalized,
  title={\href{https://ui.adsabs.harvard.edu/link_gateway/2011GReGr..43.1061J/doi:10.1007/s10714-010-1024-2}{Generalized second law of thermodynamics for a phantom energy accreting BTZ black hole}},
  author={Jamil, Mubasher and Akbar, M},
  journal={General Relativity and Gravitation},
  volume={43},
  pages={1061--1068},
  year={2011},
  publisher={Springer}
}

@article{dutta2019dark,
  title={\href{https://doi.org/10.1088/0253-6102/71/2/209}{Dark Energy Accretion onto Van der Waal’s Black Hole}},
  author={Dutta, Sandip and Biswas, Ritabrata},
  journal={Communications in Theoretical Physics},
  volume={71},
  number={2},
  pages={209},
  year={2019},
  publisher={IOP Publishing}
}

\end{document}